\documentclass{SciPost}

\hypersetup{
    colorlinks,
    linkcolor={red!50!black},
    citecolor={blue!50!black},
    urlcolor={blue!80!black}
}

\usepackage[bitstream-charter]{mathdesign}
\DeclareSymbolFont{usualmathcal}{OMS}{cmsy}{m}{n}
\DeclareSymbolFontAlphabet{\mathcal}{usualmathcal}

\fancypagestyle{SPstyle}{
\fancyhf{}
\lhead{\colorbox{scipostblue}{\bf \color{white} ~SciPost Physics }}
\rhead{{\bf \color{scipostdeepblue} ~Submission }}

\fancyfoot[C]{\textbf{\thepage}}
}

\usepackage[utf8]{inputenc}
\usepackage{amsmath}
\usepackage{amsthm}
\usepackage{mathtools}
\numberwithin{equation}{section}
\usepackage{slashed}
\usepackage{braket}
\usepackage{enumitem}
\usepackage{tabu}
\usepackage{caption}
\usepackage[svgnames]{xcolor}
\hypersetup{colorlinks,citecolor=DarkGreen,linkcolor=FireBrick,urlcolor=FireBrick,linktocpage,unicode,psdextra}
\usepackage{cite}
\usepackage{graphicx}
\usepackage{comment}
\usepackage{appendix}
\usepackage{tikz}
\usepackage{tikz-cd}
\usetikzlibrary{calc,topaths,decorations,decorations.pathmorphing,arrows,decorations.markings,cd}
\tikzset{
->-/.style args={#1rotate#2}{decoration={markings, mark=at position #1 with {\arrow[scale=1.5,rotate = #2 ]{stealth}}}, postaction={decorate}}
}

\tikzset{line/.style={line width=0.25mm},
curve/.style={line,smooth,tension=1},
->-/.style={decoration={
  markings,
  mark=at position #1 with {\arrow[>=stealth]{>}}},postaction={decorate}},
-<-/.style={decoration={
  markings,
  mark=at position #1 with {\arrow[>=stealth]{<}}},postaction={decorate}},
}

\usepackage[bottom]{footmisc}

\usepackage{courier}
\usepackage{subfig}
\usepackage{tcolorbox} 
\usepackage[normalem]{ulem}

\usepackage{mdframed}

\renewenvironment{figure}[1][]{
  \begin{originalfigure}[#1]
    \begin{mdframed}[linecolor=black!0,backgroundcolor=black!1]
}{
    \end{mdframed}
  \end{originalfigure}
}

\makeatletter
\renewcommand\section{\@startsection {section}{1}{\z@}%
                                 {-3.5ex \@plus -1ex \@minus -.2ex}
                                   {2.3ex \@plus.2ex}%
                                   {\normalfont\large\bfseries}}
\renewcommand\subsection{\@startsection{subsection}{2}{\z@}%
                                   {-3.25ex\@plus -1ex \@minus -.2ex}%
                                     {1.5ex \@plus .2ex}%
                                     {\normalfont\bfseries}}
\renewcommand\subsubsection{\@startsection{subsubsection}{3}{\z@}%
                                   {-3.25ex\@plus -1ex \@minus -.2ex}%
                                     {1.5ex \@plus .2ex}%
                                     {\normalfont\itshape}}
\makeatother

\def\cE{\varepsilon}

\def\cH{\mathcal{H}}
\def\cA{\mathcal{A}}

\def\cM{\mathcal{M}}

\def\cT{\mathcal{T}}

\def\beq{\begin{equation}}
\def\eeq{\end{equation}}
\def\barray{\begin{eqnarray}}
\def\earray{\end{eqnarray}}

\theoremstyle{definition}

\theoremstyle{definition}

\newtcolorbox{importantbox}{
  colback=blue!5!white,
  colframe=blue!75!black,
  fonttitle=\bfseries,
  title=Important 
}

\newcommand{\id}{\mathbb{1}}

\newcommand{\M}{\mathcal{M}}
\newcommand{\N}{\mathcal{N}}

\theoremstyle{definition}

\IfFileExists{ying}{
  \usetikzlibrary{external}
  \tikzexternalize[prefix=./figures/]
}

\usetikzlibrary{calc,shapes.geometric}
\pgfmathparse{atan2(0,1)}
\ifdim\pgfmathresult pt=0pt
  \tikzset{declare function={atanXY(\x,\y)=atan2(\y,\x);atanYX(\y,\x)=atan2(\y,\x);}}
\else                 
  \tikzset{declare function={atanXY(\x,\y)=atan2(\x,\y);atanYX(\y,\x)=atan2(\x,\y);}}
\fi

\begin{document}

\pagestyle{SPstyle}

\begin{center}{\Large \textbf{\color{scipostdeepblue}{
Information Loss in Generalized Symmetry Breaking
}}}\end{center}

\begin{center}\textbf{
J. Molina-Vilaplana\textsuperscript{1$\star$},
G. Sierra\textsuperscript{2} and
H-C. Zhang\textsuperscript{3$\dagger$}
}\end{center}

\begin{center}
{\bf 1} Universidad Polit\'ecnica de Cartagena, Cartagena, Spain
\\
{\bf 2} Instituto de F\'{\i}sica Te\'orica, UAM-CSIC
\\
{\bf 3} Asia Pacific Center for Theoretical Physics
\\[\baselineskip]
$\star$ \href{mailto:email1}{\small javi.molina@upct.es}
\end{center}

\section*{\color{scipostdeepblue}{Abstract}}
\textbf{\boldmath{%
We present an algebraic and information-theoretic framework for the breaking of generalized, non-invertible symmetries in $(2+1)$ dimensions. Such patterns are modeled as inclusions of finite-dimensional $C^*$-algebras equipped with conditional expectations, built upon a precise dictionary with anyon condensation in topological phases of matter. The conditional expectations are quantum channels that coarse-grain observables of the parent phase onto the symmetry-reduced condensed phase; their index --- a Watatani index equal to the quantum dimension of the condensate --- bounds, through its logarithm, the relative entropy between a state and its condensed image. This relative entropy serves as an entropic order parameter quantifying the information lost in the symmetry-reduction transition. We illustrate the framework with explicit examples: the toric code, abelian groups $Z_N$, and the representation category Rep$(S_3)$. Our results strengthen the connections between operator algebras and quantum information in the study of generalized symmetries.
}}

\vspace{\baselineskip}



\vspace{10pt}
\noindent\rule{\textwidth}{1pt}
\tableofcontents
\noindent\rule{\textwidth}{1pt}
\vspace{10pt}

\section{Introduction}
Recent advances in the study of generalized global symmetries, including the so-called non-invertible symmetries, emphasize the role of extended topological operators that implement such symmetries. In quantum field theory (QFT), these generalized global symmetries are realized through topological defect operators whose composition rules are generally governed by fusion categories \cite{Gaiotto:2014}. These symmetries impose stringent constraints on correlation functions and dualities and give rise to novel phases and topological orders \cite{McGreevy:2022, Bhardwaj:2023, Benedetti:2026}.
\medskip

In QFT, a global symmetry usually involves an ordinary compact group acting on the algebra of fields. In such cases, patterns of symmetry breaking and the resulting phases are classified by subgroups of the original symmetry group. However, in theories exhibiting topological order and non abelian anyonic excitations, the relevant statistics are governed by the braid group, and the symmetries are more naturally described by the mathematical framework of fusion categories. These structures capture what is now referred to as categorical or non-invertible symmetries, extending the notion of symmetry beyond group actions. {For these kind of generalized symmetries, the role of a subgroup is instead played by a \emph{condensable (Frobenius) algebra}, whose action organizes the possible patterns of symmetry reduction \cite{Perez-Lona:2023djo, Diatlyk:2023fwf}. Recent work extends the Landau paradigm to such symmetries through the SymTFT \cite{Bhardwaj:2023}, classifying gapped and gapless phases \cite{Bhardwaj:2023idu, Aksoy:2025, Bhardwaj:2024qiv, Bhardwaj:2025piv, Inamura:2025cum, Ji:2019jhk, Chatterjee:2022tyg, Bhardwaj:2023bbf, Wen:2023otf, Bhardwaj:2024qrf, Bhardwaj:2025jtf, Wen:2025thg}. In $(2+1)$-dimensions these reduction patterns are realized physically by \emph{anyon condensation}, the mechanism by which a set of bosonic anyons condenses into a new and reduced topological order \cite{Burnell:2018, Bischoff:2018juy} --- a transition between topological orders that lies outside the Landau paradigm of local order parameters. Throughout, “symmetry breaking” should be understood as a generalized symmetry reduction induced by anyon condensation, rather than as spontaneous symmetry breaking in the Landau sense.}
\medskip

{Here we develop an operator-algebraic and information-theoretic framework for these transitions. In this context, a symmetry reduction is modeled as an inclusion of finite-dimensional $C^*$-algebras equipped with a \emph{conditional expectation} --- coarse-graining quantum channels that projects observables from the parent phase onto the condensed one --- in precise correspondence with anyon condensation. The cost of the reduction is controlled by the index of this conditional expectation, a Watatani index which we show equals the quantum dimension of the condensate.} 
\medskip

{As an order parameter we take the quantum relative entropy between a state and its condensed image under the channel. This \emph{entropic order parameter} quantifies the information lost in the symmetry-reduction transition and obeys a universal bound given by the logarithm of the index, that is, the quantum dimension of the condensate \cite{Casini:2020, Casini:2022}. Related entropic measures have appeared in \cite{Molina-Vilaplana:2024, AliAhmad:2025, Benini:2025lav, Zhang:2025}. Relative entropy itself is the canonical tool for state distinguishability and channel-induced information loss \cite{Araki2, Araki:1976, Petz:2004, Petz:2008}.} {An entropic order parameter is especially suited for this context. Topological phases admit no local order parameters being diagnosed instead by the expectation values and mutual statistics of \emph{non-local} operators --- the loops, strings and topological defects consistent with their symmetry. Being a state-distinguishability measure intrinsic to the operator algebra, the relative entropy captures directly how the statistics of these non-local operators change between a state and its condensed image, while its monotonicity under restriction to the condensed subalgebra makes it a faithful measure of the information lost. This is what qualifies it as an order parameter for these transitions.}
\medskip

{It is worth stating at the outset what this formulation is for, and what
the resulting quantity measures. Our aim is to treat condensation as a map on
states rather than as a relabelling of anyon types, so that the cost of the
transition can be quantified, which requires the quantum channel canonically attached to the condensable algebra. This is what a conditional expectation associated with an algebra inclusion provides. The gain is not terminological: such a conditional expectation carries an index, and that index is the quantum dimension of the condensate. The entropic quantity this yields has a direct operational meaning, which we make precise in Sec.~\ref{sec:operational}: it is the exponential rate at which measurements
restricted to the observables surviving the condensation fail to distinguish a
state of the parent phase from its condensed reconstruction, so that the
quantum dimension of the condensate bounds the number of measurements required.
It is determined, as we show in Sec.~\ref{sec:examples}, by the expectation
values of the extended operators whose statistics the condensation erases.}
\medskip

{We make the construction fully explicit in three examples. For the abelian groups $\mathbb{Z}_N$ the conditional expectation is the Haar average over the group, i.e.\ the canonical projection onto a fixed-point subalgebra of index $|G|$; for the toric code we also realize the condensation channel directly on the lattice operator algebra in terms of Pauli string operators; and $\mathrm{Rep}(S_3)$ provides a genuinely non-invertible example. We further show that dualities of the parent theory organize condensations into equivalence classes. The framework thereby links operator algebras, tensor categories, and quantum information in the study of generalized symmetries.}

\smallskip
{\noindent\textbf{Outline.} Section~\ref{sec:anyon_EoP} formulates condensation as an inclusion of finite-dimensional $C^*$-algebras with a conditional expectation, expresses its index through the inclusion matrix and introduces the relative-entropy order parameter. Section~\ref{sec:examples} works out the abelian group $\mathbb{Z}_N$, the toric-code (with an explicit Pauli-operator lattice realization),  and $\mathrm{Rep}(S_3)$ examples. Two appendices collect properties of the conditional expectation (A), and the duality/equivalence of condensations (B).}

\section{Anyon Condensation and Entropic Order parameters}
\label{sec:anyon_EoP}
In this work, we study  patterns of symmetry {reduction} in $(2+1)$-dimensional theories with generalized symmetries.{We address this using} the language of topological symmetry breaking and anyon condensation, whose main mathematical description is given in terms of condensable Frobenius algebras \cite{Bais:2002, Burnell:2011, Burnell:2012,Aksoy:2025}. {Remarkably, inclusions of operator algebras provide a natural host for such generalized symmetries \cite{Longo:1994xe, Kong:2014,Kodiyalam:2001,Yu:2025}.} 

\medskip

{In our approach, the mathematical setting underlying anyon condensation is an inclusion $\N\subset\M$ of finite-dimensional $C^{*}$-algebras together with a projection map, known as conditional expectation $\varepsilon:\M\to\N$ of finite
index. When $\N$ and $\M$ are factors this is the classical setting of subfactor theory. The algebras relevant to us --- the charge algebras of anyonic superselection sectors --- are not factors; in fact $\N\subset\M$ is a multi-matrix inclusion. This does not place them outside the scope of Jones-type index theory: an index for von Neumann algebras with finite-dimensional centres was introduced by Teruya~\cite{Teruya1992}, and the general analysis of multi-factor inclusions was carried out by Giorgetti and
Longo~\cite{Giorgetti_2018}. We work instead with the index of a conditional expectation in the sense of Watatani~\cite{Watatani:1990}, which contains the Jones index~\cite{Jones:1983} as its factor special case. We do so for convenience rather than out of necessity. Watatani's definition proceeds through a quasi-basis, which in our setting is built directly from the branching data of the condensation and makes the index computable from the explicit form of the conditional expectation. 
For the finite-dimensional algebras used throughout, this is a choice of formulation and not of content. We return to this point at the end of Sec.~\ref{sec:2.2}}.
\medskip

In the physical context, we are considering a quantum mechanical system that is described by an algebra $\M$ of operators, and the dynamical variables (or observables) correspond to the self-adjoint elements. We also assume that the level of observation does not allow to access the expectation value of all observables, but only a "small" part of them represented by the subalgebra $\N$ of the full algebra $\M$.

\begin{figure}[t]
  \centering
  \includegraphics[width=0.8\textwidth]{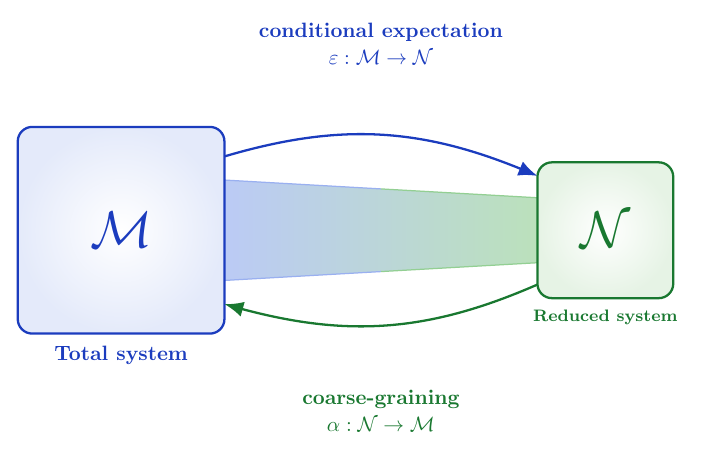}
  \caption{The conditional expectation $\varepsilon:\mathcal{M}\to\mathcal{N}$
  coarse-grains the observables of the total system $\mathcal{M}$ onto those of
  the reduced system $\mathcal{N}$, while the lift $\alpha:\mathcal{N}\to\mathcal{M}$
  embeds the reduced algebra back into the parent one. Together they encode the
  inclusion $\mathcal{N}\subset\mathcal{M}$.}
  \label{fig:MN}
\end{figure}

{A central notion is} that of a \emph{conditional expectation} $\cE: \mathcal{M} \to \mathcal{N}$. It is a unital, completely positive map that acts as a projection from $\M$ onto $\N$. A key result states that for systems with finite-dimensional Hilbert space $\cH$, a  conditional expectation $\cE$  can be expressed as:

\beq
\cE(m) = \sum_{i=1}^N \mathbb{K}_i\,  m\, \mathbb{K}^{\dagger}_i
\label{kraus}
\eeq
where $m \in \M$, and the $\mathbb{K}_i$ are the so-called \emph{Kraus operators}. In this form, a conditional expectation appears as a specific example of a quantum channel. Complementary to this is the concept of \emph{coarse-graining}. This is a positive linear embedding map or quantum channel $\alpha : \N \to \M$ that also provides partial information of the total quantum system $\M$ \emph{cf.} Fig.~\ref{fig:MN}.
\medskip

It is also necessary to introduce the notion of the \emph{index} $\lambda$ of {the conditional expectation $\cE:\M\to\N$}, a quantity that links inclusions of {operator} algebras to the concrete realizations of quantum channels in quantum information theory. {Following Watatani \cite{Watatani:1990}, which generalizes the Jones index \cite{Jones:1983} from factor inclusions to arbitrary unital $C^*$-inclusions, the index is defined through a \emph{quasi-basis} for $\cE$: a finite family $\{(u_i,v_i)\}_{i=1}^{n}\subset\M\times\M$ such that}
{\begin{equation}
m=\sum_{i} u_i\,\cE(v_i\, m)=\sum_{i}\cE(m\, u_i)\,v_i\, ,\qquad \forall\, m\in\M\, .
\label{eq:quasibasis}
\end{equation}}
{\noindent Whenever such a family exists --- always the case in our finite-dimensional setting --- the index defined by}
{\begin{equation}
\sum_{i} u_i\, v_i= \mathrm{Index}(\cE)\, \mathbb{I}_{\cM}
\label{eq:watataniindex}
\end{equation}}
{\noindent is independent of the chosen quasi-basis and belongs to the centre $Z(\M)$. For the conditional expectation we use throughout, this element is a positive scalar multiple of the identity, which we denote $\lambda$.} {If no conditional expectation exists one sets $\lambda=\infty$. The index measures the relative ``size'' of $\M$ over $\N$. It equals $|G|$ for a finite group acting with fixed-point algebra, and, for inclusions governed by a fusion category, evaluates to the total quantum dimension $\mathcal{D}^{2}$ \cite{Watatani:1990}.}
\medskip

\subsection{Charge algebras in anyon condensation}
{We consider a system of anyons carrying a generalized, non-invertible symmetry, with a Hamiltonian tuned so that a bosonic anyon $a$ condenses. Then, a new vacuum forms in which the charge $a$ becomes indistinguishable, reducing the symmetry encoded in the excitation spectrum. Before condensation the superselection sectors --- the anyon types --- form an algebra $\cA$, the anyon species of a modular tensor category, with fusion rules set by the tensor products of its objects. Condensation reduces $\cA$ to a subalgebra $\mathcal{T}$ associated to the fusion category of the surviving excitations. As a result, some operators of $\cA$ become decomposable over $\mathcal{T}$, and some that were inequivalent in $\cA$ are identified in the {symmetry-reduced} phase.}
\medskip

The condensation process and the subsequent reduction in the amount of symmetry of the system are described in terms of an algebraic object known as Frobenius or condensable algebra $\mathbb{A}$. The condensable algebra is the sum of simple objects of the uncondensed category with associated algebra $\cA$. From an algebraic operator perspective, a condensable algebra object can be thought of as a generalized analog of a subgroup, and the invariant subspace {corresponds to different patterns of symmetry reduction in the system under consideration} \cite{Yu:2025}. As will be shown, $\mathbb{A}$ defines a conditional expectation that acts as a projector from the pre-condensed algebra to the post-condensed smaller subalgebra $\varepsilon_{\mathbb{A}}: \mathcal{A}\to \mathcal{T}$, which amounts to the inclusion {$\mathcal{T}\subset \mathcal{A}$ together with the conditional expectation $\varepsilon_{\mathbb{A}}$}. That is, we decompose the larger algebra $\cA$ into elements of $\cT$. In the formalism of anyon condensation developed in ~\cite{Bais1,Bais2}, a condensation algebra is implemented in terms of the so-called \emph{restriction map}, which imposes a set of conditions that relate the particle excitations in $\cA$ to those in the condensed phase $\cT$. \footnote{In this work, we focus solely on determining the structure of $\cT$, given the original algebra $\cA$ and basic information about the condensate, such as the identification of the bosonic particles that condense. In this sense, we do not address the full topological phase transition from $\cA$ to $\mathcal{U}$, where $\mathcal{U}$ represents the algebra of non-confined charges after condensation \cite{german2016} instead concentrate solely on the symmetry reduction from $\cA$ to $\cT$.}.
\medskip

\subsection{Anyon Condensation and index of algebra inclusions}
\label{sec:2.2}
Following \cite{Bais1, Bais2, german2016} we determine the algebra $\cT$, characterized by particle labels $\phi,r,s,t,..$, along with a set of branching (or restriction) coefficients $n_a^t$. These coefficients describe how the excitations of the original uncondensed algebra $\cA$ decompose when certain bosonic sectors $\gamma_i$  condense to form a new vacuum sector $\phi$  in $\cT$:
\begin{equation}
    \phi = 1 + \gamma_1 + \cdots + \gamma_n\, .
\end{equation}
The branching coefficients implement the so-called restriction map, which characterizes the new phase:
\begin{equation}
	a\to \sum_{t \in \cT} n^t_a\,  t\, \qquad \forall a \in \cA\,  \qquad \text{(restriction)}
    \label{eq:definerestriction}
\end{equation} 
where $n_{a}^{t}\in \mathbb{Z}_{\geq 0}$. It is assumed that $n^t_a=n^{\bar{t}}_{\bar{a}}$, where bars denote antiparticles. If more than one particle $t$ appears on the right-hand side of \eqref{eq:definerestriction}, we say that ``$a$ splits into $\sum_{t} n_{a}^{t} t$''. The condensed particles satisfy the additional condition that $n^{\phi}_{\gamma_i}\neq 0$, indicating that the $\gamma_i$ sector contributes to the vacuum $\phi$ in the condensed phase $\cT$.
\medskip

In this setting, one can interpret $a$ as an operator of the algebra $\cA$, which may decompose into smaller modules when the action is restricted to a subalgebra $\cT \subset \cA$. The same branching coefficients also define the adjoint of the restriction map, commonly referred to as the lift map
\begin{equation}
	t \to \sum_{a \in \cA} n^t_a\,  a\, \qquad \forall t \in \cT\,  \qquad\text{(lift)}
\end{equation}

Consistency conditions on the branching coefficients \cite{german2016} impose two constraints on the quantum dimensions of particle excitations defined by algebras $\cA$ and $\cT$ 
\begin{subequations}
\label{eq:qtdim}
\begin{eqnarray}
\label{eq.qtdimAT}
d_a&=&\sum_{t\in\cT} n_a^t d_t,\quad\forall~a\in\cA,\\
\label{eq.qtdimTA}
d_t&=&\frac{1}{q}\sum_{a\in\cA} n^t_a d_a,\quad\forall~t\in\cT,
\end{eqnarray}
\end{subequations}
where $q$ denotes the quantum dimension of the condensed vacuum, given by 
\begin{equation}
    q:=\sum_{a \in \cA} n^\phi_a d_a = D_\cA^2/D_\cT^2\, ,
    \label{eq:define_q}
\end{equation}
 where 
 \begin{align}
     D^2_\cA =\sum_{a \in \cA}\, d_a^2\, , \quad  D^2_\cT =\sum_{t \in \cT}\, d_t^2\, ,
 \end{align}
 are the total (Perron-Frobenius) quantum dimensions of $\cA$ and $\cT$ respectively. {Inclusions of finite-dimensional $C^*$-algebras with their conditional expectations offer a natural framework for describing anyon condensation: in this language, the (Watatani) index} $\lambda$  corresponds to the quantum dimension of the condensate $q$. {To see this, it is convenient to phrase this index directly in terms of the branching data, through the \emph{inclusion matrix} of $\cT \subset \cA$,
\begin{equation}
\label{eq:inclusionmatrix}
\Lambda_{at} := n^t_a\, , \qquad a \in \cA,\ t \in \cT\, ,
\end{equation}
which is nothing but the biadjacency matrix of the bipartite  inclusion
graph (Bratteli diagram) displayed in the examples below \footnote{In the subfactor case this object is known as the principal graph, constructed from the Jones tower.}. Writing $\vec d_{\cA} = (d_a)_{a\in\cA}$ and $\vec d_{\cT} = (d_t)_{t\in\cT}$ for the quantum-dimension vectors, the consistency conditions \eqref{eq.qtdimAT}--\eqref{eq.qtdimTA} are precisely the statements
\begin{equation}
\label{eq:PFrelations}
\Lambda\, \vec d_{\cT} = \vec d_{\cA}\, , \qquad \Lambda^{\!\top} \vec d_{\cA} = q\, \vec d_{\cT}\, ,
\end{equation}
and hence $\Lambda^{\!\top}\Lambda\, \vec d_{\cT} = q\, \vec d_{\cT}$. Since $\vec d_{\cT}$ has strictly positive entries, the Perron--Frobenius theorem identifies $q$ with the spectral radius of $\Lambda^{\!\top}\Lambda$, so that the index is the squared norm of the inclusion matrix,
\begin{equation}
\label{eq:indexnorm}
\lambda =  q = \Vert \Lambda \Vert^2 
\end{equation}
with the quantum dimensions playing the role of the Perron--Frobenius weights of the inclusion. Reading off the vacuum component $t=\phi$ of the second relation in \eqref{eq:PFrelations}, with $d_\phi = 1$, recovers the explicit value $\lambda = \sum_a n^\phi_a d_a$ quoted above.} 
\medskip

{The equality between the quasi-basis definition \eqref{eq:watataniindex} of the Watatani index and the Perron--Frobenius expression \eqref{eq:indexnorm} can be read directly off the branching data. Since $\cA$ is abelian, the index is a central element, $\mathrm{Index}(\varepsilon_{\mathbb A})\mathbb{I}_{\cA}=\sum_{a \in \cA}\big(\mathrm{Index}\big)_a\,\Pi_a$, where $\Pi_a$ is a central projector on sector $a$. Then, it suffices to show that all of its components coincide. The relevant quasi-basis is not labeled by the parent sectors that branch to the vacuum, but by the two-step process $a\to t\leftarrow b$ --- restriction from $\cA$ to a condensed sector $t$, followed by the adjoint lift back to $\cA$ --- whose adjacency matrix is $\Lambda\Lambda^{\!\top}$. Concretely, the explicit quasi-basis $\{(u_{at},v_{at})\}$ built in Section~\ref{sec:cond_exp} yields}
{\begin{equation}
\big(\mathrm{Index}(\varepsilon_{\mathbb A})\big)_a
=\frac{1}{d_a}\sum_{t\in\cT}\sum_{b\in\cA} n^t_a\, n^t_b\, d_b
=\frac{\big(\Lambda\Lambda^{\!\top}\vec d_{\cA}\big)_a}{d_a}\, .
\end{equation}}
{\noindent For the canonical conditional expectation the quantum-dimension vector $\vec d_{\cA}$ is the Perron--Frobenius eigenvector of this two-step adjacency,}
{\begin{equation}
\Lambda\Lambda^{\!\top}\,\vec d_{\cA}=q\,\vec d_{\cA}\, ,
\end{equation}}
{\noindent so that $\big(\mathrm{Index}\big)_a=q$ for every $a$. Thus, the central element collapses to the scalar $\mathrm{Index}(\varepsilon_{\mathbb A})=q$. As $\Lambda\Lambda^{\!\top}$ and $\Lambda^{\!\top}\Lambda$ share their nonzero spectrum, $q=\rho(\Lambda\Lambda^{\!\top})=\Vert\Lambda\Vert^2$, recovering \eqref{eq:indexnorm}. This expression reproduces the standard values \cite{Watatani:1990}; for an invertible (group) symmetry this returns $|G|$, and for a condensation of all the "a"-charges the total categorical dimension $\sum_a d_a^2=\mathcal D^2$. This scalar identity holds for the conditional expectation realized by our construction; for a generic one the Watatani index is a non-scalar element of the centre of $\cA$.} 
\medskip

{The expressions above admit a reading in terms of the theory of multi-factor inclusions. The authors of Ref.~\cite{Giorgetti_2018}, extending earlier work of Ref.~\cite{Teruya1992}, characterise the minimal conditional expectation for inclusions of von Neumann algebras with finite-dimensional centres through a canonical state on the relative commutant which they call the spherical state, and show that the minimal index is the squared norm of a matrix dimension, as was already known for multi-matrix inclusions. The two relations~\eqref{eq:PFrelations} play distinct roles in this comparison. The first, $\Lambda\,\vec d_{\cT}=\vec d_{\cA}$, is the statement that each sector of $\cA$ decomposes into sectors of $\cT$ with
multiplicities $\Lambda_{at}$, and holds for any such inclusion. The second, $\Lambda^{\!\top}\vec d_{\cA}=q\,\vec d_{\cT}$, is the Markov condition on the weights $\vec d$, and it is what makes $\vec d_{\cA}$ the Perron--Frobenius eigenvector of $\Lambda\Lambda^{\!\top}$
and $\lambda=\Vert\Lambda\Vert^{2}$ a candidate for the minimal index. Since the Watatani index is attached to a choice of conditional expectation, we do not claim its minimality  in general. In each of the examples of Sec.~\ref{sec:examples} both relations
in~\eqref{eq:PFrelations} are satisfied with integral quantum dimensions, so
that $\lambda=\Vert\Lambda\Vert^{2}$ coincides with the value that the minimal
index of Ref.~\cite{Giorgetti_2018} takes for a multi-matrix inclusion with
inclusion matrix $\Lambda$. Whether the identification persists beyond the cases
treated here we leave open. We note, finally, that the index bound on the entropic order parameter is insensitive to this question, since it holds for any conditional expectation of finite index.
}

\subsection{Conditional expectation for the restriction and lifting maps}
\label{sec:cond_exp}
 Here, from a quantum information perspective, we describe the {symmetry reduction transition} of a generalized symmetry and its inversion using the language of anyon condensation. {In operator-algebraic terms, anyon condensation can be described in terms of a conditional expectation which implements a projection onto invariant observables under the condensation map.}
\medskip

Our framework is thus based on this unifying perspective, making use of the \emph{restriction} and \emph{lifting} maps to build the quantum channels (conditional expectations) yielding density operators corresponding to different patterns of (generalized) symmetry reduction-as 
        \begin{align*}
            {\rm Condensable\, Algebra\, \mathbb{A}}\to n^t_a\implies \varepsilon_{\mathbb{A}}: \rho \in \cA \longrightarrow \rho_{\mathbb{A}} \in \mathcal{T}
        \end{align*}
The density operator $\rho_{\mathbb{A}} =\varepsilon_{\mathbb{A}}(\rho)$ represents an averaged version of the density operator $\rho$ w.r.t. {condensable algebra} $\mathbb{A}$. The coarse-graining $\alpha$ that implements the lifting map, expresses the {averaged} density operator in terms of the original algebra $\mathcal{A}$
         \begin{align*}
         \alpha: \rho_{\mathbb{A}} \in \mathcal{T} \longrightarrow \tilde{\rho} \in \mathcal{A}\, .
        \end{align*}

\begin{figure}[t]
  \centering
  \includegraphics[width=0.8\textwidth]{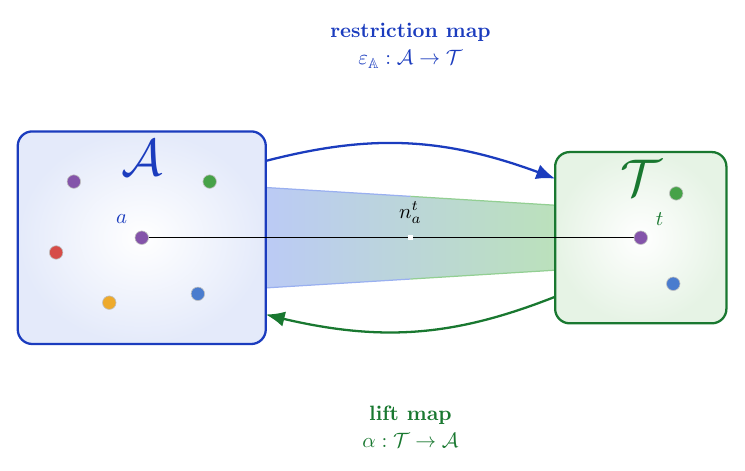}
  \caption{Restriction and lift maps between the parent charge algebra $\mathcal{A}$
  and the condensed algebra $\mathcal{T}$. The restriction
  $\varepsilon_{\mathbb{A}}:\mathcal{A}\to\mathcal{T}$ projects the anyon content of
  the uncondensed phase onto the sectors surviving condensation, with branching
  coefficients $n^t_a$; the lift $\alpha:\mathcal{T}\to\mathcal{A}$ embeds them back.
  Condensation usually reduces the number of distinct species, so $\mathcal{T}$ carries fewer
  charges than $\mathcal{A}$.}
  \label{fig:AT}
\end{figure}

{The physical setup we consider is as follows: We have a 2D quantum system with anyonic excitations described by algebra $\cA$ where the observable topological charges are labeled by projectors $\Pi_a$. { Here $\cA$ is the abelian algebra generated by these charge-sector projectors, i.e.\ the global superselection data of the phase, rather than the full local operator algebra of a lattice model; a concrete lattice realization is given for the toric code in Section~\ref{sec:toric}.} A quantum state $\rho$ assigns probabilities $p_a$ to finding charge $a$. The condensation of certain bosonic anyons redefines the vacuum in such a way that, after condensation, the reduced algebra $\cT$ describes the surviving excitations. The conditional expectation $\varepsilon_{\mathbb{A}}: \cA \to \cT$ implements the physical process of "forgetting" which condensed anyons are present. The lifted state $\tilde \rho = \alpha \circ \varepsilon_{\mathbb{A}}(\rho)$ represents our best reconstruction of the original state using only post-condensation observables, \emph{cf} Fig.~\ref{fig:AT}.}
\medskip

In this sense, we have discussed above how $C^*$- algebra inclusion $\N \subset \M$ defined by a conditional expectation $\varepsilon: \cM \rightarrow \N$ can be implemented via Kraus operators, as shown in Eq.~\eqref{kraus}. Turning to the inclusion $\cT \subset \cA$, here we consider how to build the corresponding conditional expectation in terms of Kraus operators $\mathbb{K}_a$ as,
\begin{align}
    \varepsilon_{\mathbb{A}}(\rho)=\sum_{a \in \cA} \mathbb{K}_a\, \rho\, \mathbb{K}^{\dagger}_a\, .
\end{align} 
\medskip

First, we note that the structure of the state space (an operator space) of an anyonic system prohibits superpositions between different topological charges due to anyonic superselection rules \cite{Kitaev:2003, Bonderson:2008}. These superselection rules require that an anyonic state be represented by a density operator that is a direct sum of sectors $\bigoplus_a \rho_a$. As a result, we consider an Abelian algebra $\cA$ representing the uncondensed phase. For this algebra, the fusion rules of the non-local symmetry operators can be diagonalized in terms of a set of projectors $\Pi_a$ on the states of $\cA$  with charge $a$ in such a way that in the standard representation of the algebra, a general state is given by the density matrix
\begin{equation}
    \rho = \sum_{a \in \cA}\, p_a\,  \Pi_a \, , 
    \label{eq:rho_ss}
\end{equation}
where projectors $\Pi_a$  satisfy the following conditions
\beq
\Pi_a^\dagger = \Pi_a, \quad 
\Pi_a \Pi_b = \delta_{ab} \Pi_a
\eeq
and  $\{ p_a \}$ are the corresponding probabilities $p_a \equiv {\rm tr} (\rho\, \Pi_a)$ such that  $\sum_a p_a =1$.
\footnote{
As an example, at infinite temperature $\beta \to 0$ one has that \cite{Benedetti:2024dku}
\begin{equation}
    p_a=\frac{N_a\, d_a}{\sum_{b \in \cA}N_b\, d_b} \, ,
    \label{eq:probs}
\end{equation}
with $N_a$ referring to the multiplicity of the irrep $a$ and $d_a$ the corresponding quantum dimensions. }
\medskip

To construct the conditional expectation $\varepsilon_{\mathbb{A}}:\cA \to \cT$ that implements the restriction map associated with a condensable algebra $\mathbb{A}$, we define the corresponding Kraus operators using the branching coefficients $n^t_a$ as follows,

\begin{align}
        \mathbb{K}_a &=\sum_{t \in \cT}\,\sqrt{\, n^t_a\left(\frac{d_t}{d_a}\right)}\, 
        \Delta_{t,  a} \, , 
        \quad a\in \cA , \, \quad  t \in \cT\, , 
     \label{eq:kraus_op}
\end{align}
where $\Delta_{t,a}$, and its adjoint $\Delta_{a,t}= \Delta_{t,a}^\dagger$ are operators that satisfy 
\beq
\Delta_{t,a} \Pi_b = \delta_{ab} \Delta_{ta}, \quad  \Delta_{t,a} \Delta_{b, s} = \delta_{a b}  \delta_{t s}  \Pi_t  \, . 
\eeq
With this definition, the action of $\varepsilon_{\mathbb{A}}$ on $\rho$ is given by 
\begin{align}
     \varepsilon_{\mathbb{A}}(\rho)&=\sum_{a \in \cA}\, \mathbb{K}_a\, \rho\, \mathbb{K}_a^{\dagger} = \sum_{a \in \cA}\, \mathbb{K}_a\, \left( \sum_{a \in \cA}\, p_a\, \Pi_a \right)\,\mathbb{K}_a^\dagger  \nonumber \\
     &=\sum_{t,a}\, n^t_a\left(\frac{d_t}{d_a}\right)\, p_a\, \Pi_t=\sum_{t\in \cT}\, p_t\, \Pi_t\, 
\end{align}
where the probabilities in the condensed phase are
\begin{align}\label{eq:ptcond}
     p_t= \sum_{a\in \cA}\, n^t_a\left(\frac{d_t}{d_a}\right)\, p_a\, .
\end{align}

{Physically, the conditional expectation $\varepsilon_{\mathbb{A}}$ {is the symmetry-averaging map onto the condensed charge algebra: it retains the information of} which post-condensation charge $t$ is present, but {discards} which pre-condensation charge $a$ produced it (when multiple $a$'s branch to the same $t$). The Kraus operators $\mathbb{K}_a$ in Eq. \eqref{eq:kraus_op} implement this coarse-graining.}

So defined, the conditional expectation $\cE_{\mathbb{A}}$ is trace preserving, idempotent (so acts as a projector) and fulfills the bimodule property. The proofs of the latter are presented in Appendix A. The trace-preserving property of the conditional expectation $\varepsilon_{\mathbb{A}}$ can be easily verified as follows:

\begin{align}
    {\rm tr }\,  \varepsilon_{\mathbb{A}}(\rho)=\sum_{t \in \cT}\, p_t = \sum_{a \in \cA} \frac{1}{d_a}\, p_a\, \sum_{t \in \cT}\, n^t_a\, d_t = \sum_{a \in \cA}\, p_a = {\rm tr}\,  \rho \, , 
\end{align}
where in the final step we used the quantum dimension identity \eqref{eq.qtdimAT}. 
Furthermore, the Kraus operator condition is satisfied:
\begin{align}
    \sum_{a \in \cA}\, \mathbb{K}_a^{\dagger}\, \mathbb{K}_a &= \sum_{a\in \cA}\sum_{t,s \in \cT}\, \sqrt{n^t_a\, n^s_a \frac{d_t d_s}{d_a^2}}\, 
    \Delta_{a, t} \Delta_{s,a} 
    = \sum_{a\in \cA}\, \frac{1}{d_a}\, \left(\sum_{t\in \cT}n^t_a\, d_t\right)\, 
    \Pi_a 
    \\      \nonumber
    &=\sum_{a\in \cA}\,  
    \Pi_a 
    =     \mathbb{I}_{\cA}\, .
    \end{align}

{This last relation immediately allows to define the conditional expectation and its index in terms of an explicit quasi basis given by
\begin{equation}
    u_{at}=\sqrt{\lambda\, n^t_a\, \left(\frac{\, d_t}{d_a}\right)}\, \Delta_{a,t}\quad v_{at}=\sqrt{\lambda\, n^t_a\, \left(\frac{\, d_t}{d_a}\right)}\, \Delta_{t,a}\, .
\end{equation}
where 
\begin{align}
    \lambda=\sum_{a \in \cA} n^\phi_a d_a = q\, ,
\end{align}
is the {Watatani}-index of the conditional expectation. Then, noting that
\begin{equation}
    \mathbb{K}_a=\frac{1}{\sqrt{\lambda}}\, \sum_{t \in \cT}\, v_{at}\, ,\quad \mathbb{K}^{\dagger}_a=\frac{1}{\sqrt{\lambda}}\, \sum_{t \in \cT}\, u_{at}\, ,
\end{equation}
one obtains that
\begin{equation}
    \sum_{a \in \cA}\mathbb{K}_a^{\dagger}\,\mathbb{K}_a=\frac{1}{\lambda}\, \sum_{a \in \cA}\, \sum_{t \in \cT}\, u_{at}\, v_{at}\, = \mathbb{I}_{\cA} , \implies \sum_{a \in \cA}\,  \sum_{t \in \cT}\,u_{at}\, v_{at}\, = \lambda\, \mathbb{I}_{\cA}
\end{equation}
}

In addition to this, one can define the coarse-graining quantum channel $\alpha: \cT \to \cA$ that implements the lifting map through:
\begin{equation}
    \mathbb{L}_{a} =\sum_{t \in \cT}\, \sqrt{\frac{n^t_a}{\lambda}\left(\frac{d_a}{d_t}\right)}\, \Delta_{a,t} \, \quad a\in \cA\, \quad  t \in \cT\, , 
\end{equation}
 With this, the lifted density matrix $\tilde{\rho}$ is given by
\begin{align}\label{tilderho}
     \widetilde\rho&= \alpha \circ \varepsilon_{\mathbb{A}}(\rho)=\sum_{a \in \cA}\, \mathbb{L}_{a}\, \varepsilon_{\mathbb{A}}(\rho)\,  \mathbb{L}_{a}^{\dagger} =\sum_{a \in \cA}\, \mathbb{L}_{a}\, \left(\sum_{t\in \cT}\, p_t\, 
     \Pi_t \right)\,  \mathbb{L}_{a}^{\dagger} \\ \nonumber
     &= \frac{1}{\lambda}\sum_{t,a}\, n^t_a\left(\frac{d_a}{d_t}\right)\, p_t\, \Pi_a =\sum_{a\in \cA}\, \widetilde{p}_a\, \Pi_a \, ,
\end{align}
where the lifted probabilities $\tilde{p}_a$ are given by
\begin{align}
     \widetilde{p}_a&=\frac{1}{\lambda} \sum_{t\in \cT}\, n^t_a\left(\frac{d_a}{d_t}\right)\, p_t\, .
     \label{tildep}
\end{align} 

Although the operators $\mathbb{L}_{a}$
do not form a set of Kraus operators, they nonetheless preserve the trace of
$\tilde{\rho}$, that is, 
$\sum_{a \in \cA} \tilde{p}_a=1$,
as ensured by the identity \eqref{eq.qtdimTA}. 
Moreover, the operators 
$\mathbb{L}_{a}$ satisfy an interesting trace identity:

\begin{align}
    \sum_{a \in \cA} {\rm tr}\left(\mathbb{L}_a \mathbb{L}_a^{\dagger}\right) =  |\cT|\, , 
\end{align}
where $|\cT|$ denotes the number of particle types (or simple objects) in the condensed theory.
\medskip

{Summarizing, the conditional expectation $\varepsilon_{\mathbb{A}}$ is completely positive, trace preserving, unital and idempotent (Appendix~\ref{AppendixA}), hence a genuine quantum channel. The lifting map $\alpha$ is positive and trace preserving but not unital, so the operators $\mathbb{L}_a$ do not form a Kraus family; nonetheless $\alpha \circ \varepsilon_{\mathbb{A}}$ is a well-defined trace-preserving completely positive map on $\cA$. From the perspective of operator algebras, the lifting map $\alpha:\, \cT\to \cA\, $ can be viewed as a state-level analogue of the well-known $\alpha$-induction procedure \cite{Bockenhauer1, Bockenhauer2}, which lifts superselection sectors using the same branching data that enter our formulas; here we obtain its decategorified-braiding-blind counterpart, acting instead on states.}
\medskip

As a remark, we notice that Equation \eqref{tildep} can also be written as 

    \begin{align}
     \widetilde{p}_a&=\frac{1}{\lambda} \sum_{b\in \cA}\, M'_{ab}\left(\frac{d_a}{d_b}\right)\, p_b\, ,\label{tildep2}
\end{align}
with the matrix 
\beq
M'_{ab} = \sum_{t \in \cT} n^t_a n^t_b 
\label{eq:Mprime}
\eeq
defined in \cite{german2016}. {Equivalently, $M' = \Lambda\Lambda^{\!\top}$ is the adjacency matrix of the inclusion matrix \eqref{eq:inclusionmatrix} on the $\cA$ side; it shares the nonzero spectrum of $\Lambda^{\!\top}\Lambda$, so that $\Vert M'\Vert = \lambda$, and the quantum-dimension vector is its Perron--Frobenius eigenvector, $\sum_b M'_{ab}\, d_b = \lambda\, d_a$.} \footnote{As stated in \cite{german2016}, restricting the summation in this equation to the unconfined particles $u \in \mathcal{U}$, one obtains the matrix $M_{ab}=\sum_{t \in \mathcal{U}}n^t_a \, n^t_b$  that plays a central role in the construction of the MTC $\mathcal{U}$  that, as explained above, is not the purpose of our work.}
\medskip

{It is worth making explicit here the connection between $M'$, the inclusion matrix $\Lambda$ and the branching coefficients. The condensable algebra $\mathbb A=\bigoplus_{c} n^{\phi}_{c}\,c$ specifies only the column $t=\phi$ of the inclusion matrix~\eqref{eq:inclusionmatrix}, namely the decomposition of the new vacuum into sectors of $\cA$. The remaining branching coefficients are fixed by the consistency conditions of Ref.~\cite{german2016}, as we now show.}

{Besides the quantum-dimension conditions~\eqref{eq.qtdimAT}--\eqref{eq.qtdimTA}
and the reality condition $n^{t}_{a}=n^{\bar t}_{\bar a}$, the restriction
map~\eqref{eq:definerestriction} is required to be compatible with fusion:
restricting both sides of $a\times b=\sum_{c}N^{c}_{ab}\,c$ gives
\begin{equation}
  \sum_{c\in\cA} N^{c}_{ab}\, n^{t}_{c}
  \;=\;\sum_{r,s\in\cT} n^{r}_{a}n^{s}_{b}\,\widetilde N^{t}_{rs},
  \label{eq:ring-hom}
\end{equation}
where $N$ and $\widetilde N$ are the fusion coefficients of $\cA$ and $\cT$.
Setting $t=\phi$ in \eqref{eq:ring-hom} and using that the vacuum appears in
$r\times s$ precisely when $s=\bar r$, and then once, so that
$\widetilde N^{\phi}_{rs}=\delta_{s\bar r}$, the right-hand side collapses to
$\sum_{r} n^{r}_{a} n^{\bar r}_{b}$, which the reality condition turns into
$\sum_{r} n^{r}_{a} n^{r}_{\bar b}$. Replacing $b\to\bar b$ and using
$N^{c}_{a\bar b}=N^{a}_{bc}$, this is precisely the matrix $M'$
of~\eqref{eq:Mprime}:
\begin{equation}
  M'_{ab}\;=\;\sum_{t\in\cT} n^{t}_{a}n^{t}_{b}
  \;=\;\sum_{c\in\cA} n^{\phi}_{c}\,N^{a}_{bc}.
  \label{eq:Mprime-from-A}
\end{equation}
}

{The whole of $M'$ is thus determined by the condensable algebra together with
the fusion rules of the parent theory, without prior knowledge of $\cT$. Two
immediate checks follow: setting $b=1$ returns the condensable algebra,
$M'_{a1}=n^{\phi}_{a}$, and $M'_{11}=n^{\phi}_{1}=1$ expresses the
connectedness of $\mathbb A$.}

{The inclusion matrix itself is then obtained by factorising $M'$ over the
nonnegative integers, $M'=\Lambda\Lambda^{\!\top}$, with the vacuum column
prescribed and subject to the quantum-dimension
conditions~\eqref{eq.qtdimAT}--\eqref{eq.qtdimTA}, the reality condition, and
the total dimension $\mathcal D^{2}_{\cT}=\mathcal D^{2}_{\cA}/q$
of~\eqref{eq:define_q}. This is the linear-algebra problem formulated in
Ref.~\cite{german2016}. The conditions are necessary and, while sufficient in
all cases treated below, need not determine $\Lambda$ uniquely in general. That is to say, the factorization need not be unique in general. Possible factorizations are
therefore candidates that must be checked against the remaining consistency
conditions.}

\subsection{Relative Entropy as order parameter}
With the previous constructions in place, we are now in a position to entropically quantify the information loss due to symmetry breaking after the condensation process using the relative entropy. The relative entropy is a central quantity in quantum information theory. For matrix algebras, it is defined by
\begin{equation}
    S_{\M}(\rho\, \vert\, \sigma)= {\rm tr}_{\M} [\rho(\log \rho - \log \sigma)]\, \label{relative} ,
\end{equation} 
where $\textrm{tr}_{\M}$ denotes the canonical trace associated with the algebra $\M$, and $\rho$, $\sigma$ are the two density matrices associated with the two states in comparison; see \cite{Petz:2004}. Relative entropy provides a useful measure of how distinguishable two quantum states are. 
\medskip

Although the computation of this quantity in QFT is not straightforward, in \cite{Casini:2020} the authors noted that this simplifies when, as in our case, $\cA$ and $\cT$ are small algebras for which the expectation values of the nonlocal operators are known. That is, the expectation values of the projectors $\Pi_a$ in both states $\rho$ and $\tilde \rho = \alpha \circ \varepsilon_{\mathbb{A}}(\rho)$ are 
\begin{equation}
    p_a \equiv {\rm tr} (\rho\, \Pi_a)\,,\qquad \tilde p_a \equiv {\rm tr} (\tilde\rho\, \Pi_a)\;,
\end{equation} one finds that the relative entropy is given by 

\begin{equation}
    S_{\cA}(\rho\, \vert\, \alpha \circ \varepsilon_{\mathbb{A}}(\rho))={\rm tr}_{\cA} [\rho(\log \rho - \log \tilde{\rho}]\, 
    = \sum\limits_{a \in \cA}\, p_a\, \log \frac{p_a}{\tilde{p}_a} \label{relative2} . 
\end{equation} 
This quantity provides a way to compare a density matrix in the algebra $\cA$ before and after the condensation process. We stress that in our setting, this quantity characterises the transition rather than the condensed phase. It compares a state of the parent algebra with its image under the condensation channel, and is not a function of the condensed state alone. The name ``entropic order parameter'' is inherited from Refs.~\cite{Casini:2020, Magan:2020ake}, where the same construction is applied to ordinary symmetries. As there, the algebraic data fix the range of the quantity while the state fixes its value.

{It qualifies as an order parameter because $\tilde\rho = \alpha\circ\varepsilon_{\mathbb{A}}(\rho)$ differs from $\rho$ only through the expectation values of the non-local symmetry operators (the loop/charge operators diagonalized by the $\Pi_a$): it vanishes precisely on the symmetric states $\rho = \alpha\circ\varepsilon_{\mathbb{A}}(\rho)$ and is strictly positive once a charged, non-local one-point function is switched on, as made explicit in the perturbative analysis below.} 

An equivalent expression, derived using Eq.~\eqref{tildep2}, is

\begin{equation}
    S_{\cA}(\rho\, \vert\, \alpha \circ \varepsilon_{\mathbb{A}}(\rho)) = \log \lambda -H(p) - \sum\limits_{a \in \cA}\, p_a\, \log\left[\sum \limits_{b \in \cA}\, M'_{ab}\left(\frac{d_a}{d_b}\right)\, p_b\right]\;.
     \label{eq:rel_coarse}
\end{equation}
where 
$H(p)$  is the Shannon entropy of the initial probability distribution
$\{ p_a \}$, defined as
\begin{align}
    H(p) =-\sum_{a\in \cA} p_a \log p_a\, .
\end{align}

The matrix $M'_{ab}$ in our coarse-graining formula \eqref{eq:rel_coarse} coincide with the coupling matrices that appear in the $\alpha$-induction framework \cite{Bockenhauer1, Bockenhauer2}. There, such matrices encode modular invariants governing the extension of superselection sectors. In our setting, the same data control the probabilities $\tilde p_a$ of the lifted states and hence the loss of distinguishability as measured by relative entropy. Namely, for well-defined probability distributions and restriction-lifting maps, there is a bound, {given by the Watatani index}
\begin{align}
     S_{\cA}(\rho\, \vert\, \alpha \circ \varepsilon_{\mathbb{A}}(\rho)) \leq \log \lambda
     \label{bound}
\end{align}
for the maximum amount of information loss associated to a {pattern of symmetry reduction} given in terms of the index of the algebra inclusion. Bounds of this type for inclusions of type~I algebras with finite-dimensional centres, and their relation to the index, were first obtained in \cite{Casini:2020, Magan:2020ake} (see also \cite{AliAhmad:2025}). {In less abstract terms, how well can one distinguish the pre-condensation state $\rho$ from the lifted reconstruction $\alpha \circ \varepsilon_{\mathbb{A}}(\rho)$? A large value of $S_{\cA}(\rho\, \vert\, \alpha \circ \varepsilon_{\mathbb{A}}(\rho))$ indicates that condensation destroys significant information about the original state. 
This also leads to operationally define a symmetric states as those fulfilling
\begin{equation}
    \rho=\alpha \circ \varepsilon_{\mathbb{A}}(\rho)\, .
\end{equation}} 

\subsection{Operational meaning of the entropic quantity}
\label{sec:operational}

The relative entropy \eqref{relative} is not only an abstract measure of
distinguishability. In the present setting it has an exact
operational reading. Since the charge algebra $\cA$ is abelian, comparing
$\rho$ with its condensed reconstruction
$\tilde\rho=\alpha\circ\varepsilon_{\mathbb A}(\rho)$ amounts to a classical
hypothesis test between two probability distributions over superselection
sectors. Here, the Chernoff--Stein lemma applies directly as follows: given $N$
independent measurements of the charge observables of the parent phase,
the probability of failing to distinguish $\rho$ from $\tilde\rho$ decays as
\begin{equation}
  p\;\sim\;e^{-N\,S(\rho\Vert\tilde\rho)}.
  \label{eq:stein}
\end{equation}
The entropic quantity is therefore the exponential rate at which
measurements restricted to the observables that survive condensation fail
to reconstruct the state one began with. Equivalently, achieving a fixed
confidence requires $N\sim 1/S(\rho\Vert\tilde\rho)$ measurements. The bound \eqref{bound} inherits this interpretation. Since
$S(\rho\Vert\tilde\rho)\le\log\lambda$ with $\lambda=q$ the quantum dimension of
the condensate, the condensate sets the maximum attainable rate and hence
a lower bound
\begin{equation}
  N\;\gtrsim\;\frac{1}{\log q}
  \label{eq:Nbound}
\end{equation}
on the number of measurements required. An algebraic invariant of the
condensation thus controls an operational quantity: the larger the
condensate, the more sharply the parent state can be told apart from its
condensed image, and the information erased in the transition is bounded
by $\log q$ regardless of the state.%

\section{Examples}
\label{sec:examples}
In this section, we present a selection of representative examples that illustrate the ideas discussed above and show the relevance of our construction to {study patterns of symmetry reduction in systems with generalized symmetries}.  The entropic order parameter $ S_{\cA}(\rho | \alpha \circ \varepsilon_{\mathbb{A}}(\rho))$ in Eq. \eqref{relative2} quantifies the entropic distance between a state $\rho$ and its projection to the subalgebra  $\cT$ associated with the (condensable) Frobenius algebra $\mathbb{A}$ defining a symmetry reduced sector of the theory.

\subsection{Abelian group $\mathbb{Z}_N$.}
Let us first focus on the group $\mathbb{Z}_N$, for which we consider the condensable algebra,
 
\begin{align}
    \mathbb{A}=\sum\limits_{r }\, r\, 
\end{align} 
with $r \in {\rm irreps}\,  G$. In this case, $\cT=\lbrace \phi\rbrace$ with $n^{\phi}_r=1,\, \forall\, r$ and $p_{\phi}=1$. The index $\lambda$ is given by
\begin{equation}
    \lambda=|G|\, .
\end{equation}

  {In the mathematical literature, the condensable algebra $\mathbb{A}$ above is known as a \emph{Lagrangian} algebra.} In terms of branching coefficients, {a Lagrangian} condensable algebra is defined through coefficients that satisfy \emph{i)} $n^{\phi}_a \neq 0, \, \forall\, a \in \cA$, and \emph{ii)} that the quantum dimension of the condensed vacuum $q$, that is, the index $\lambda$ is
\begin{align}
    \lambda = \sum_{a \in \cA}\, n^{\phi}_a\, d_a = \mathcal{D}_{\cA}^2\, \implies \mathcal{D}_{\cT}^2=1\, .
\end{align}
Thus, the pattern of symmetry breaking results in a new vacuum $\phi$ in which all the non-trivial anyonic charges $a$ have condensed. We will see another example of Lagrangian algebra in the case of non-invertible symmetries. 
\medskip

Defining 
\begin{align}
    {\rm n}^{\phi}=\begin{pmatrix}
        n_1^{\phi} & n_2^{\phi} & \cdots & n_N^{\phi}
    \end{pmatrix} = \begin{pmatrix}
        1 & 1 & \cdots & 1
    \end{pmatrix}
\end{align}

Since all $n^{\phi}_{c}=1$ and the fusion of $\mathbb{Z}_N$ is the group
law, one has $\sum_{c}N^{a}_{bc}=1$ for every pair $(a,b)$, so that
\eqref{eq:Mprime-from-A} gives directly the $N\times N$ all-ones matrix
\begin{align}
    M'=({\rm n}^{\phi})^{\rm T}{{\rm n}^{\phi}} =\begin{pmatrix}
    1 & \cdots & 1 \\
    \vdots & \ddots & \vdots  \\
    1 & \cdots & 1  \\
    \end{pmatrix}
\end{align}
{In the notation of \eqref{eq:inclusionmatrix}, the inclusion matrix here is the single column $\Lambda=(\mathrm{n}^{\phi})^{\!\top}=(1,\dots,1)^{\!\top}$, so that $\Lambda^{\!\top}\Lambda = N = |G| = \lambda$ while $M'=\Lambda\Lambda^{\!\top}$ is the $N\times N$ all-ones matrix displayed above. The index is its single nonzero eigenvalue $\Vert\Lambda\Vert^2 = N$.}
From this matrix $M'_{rs}$ it is straightforward to see that 
\begin{align}
    \sum_ s\, M'_{rs}\,  p_s = p_{\phi} = 1\, \quad \forall r\, ,
\end{align}
which implies that
\begin{equation}
     S_{\cA}(\rho |\, \alpha \circ \varepsilon_{\mathbb{A}}(\rho)) =\log \lambda +\sum\limits_{r}\, p_r\, \log p_r = \log |G| -H(p)\, ,
     \label{casezn}
\end{equation}
a result that was first obtained in \cite{Casini:2020}. It is worth noting that the lifted density matrix $\tilde{\rho}$ is proportional to the identity 
\beq
\tilde{\rho}= \alpha \circ \varepsilon_{\mathbb{A}}(\rho)=\frac{1}{\lambda}  \sum_{r} \, \Pi_r
\eeq
which corresponds to a maximally mixed state and gives rise to the $\log |G|$ contribution in the relative entropy 
$S_{\cA}(\rho |\, \alpha \circ \varepsilon_{\mathbb{A}}(\rho))$. {This maximally mixed lift has a transparent group-theoretic origin that makes the conditional expectation fully explicit. Let $\{V_g\}_{g\in\mathbb{Z}_N}$ be the  unitaries acting on the sectors by $V_g\,\Pi_r\,V_g^{\dagger}=\Pi_{r+g}$ (indices mod $N$). They furnish a representation of $G=\mathbb{Z}_N$ on $\cA$ whose only invariants are multiples of the identity, $\mathbb{C}\,\id$, which we identify with the condensed theory $\cT$. For any state $\rho=\sum_r p_r\,\Pi_r$ the lift is then nothing but the Haar average over $G$,}
{\begin{equation}
\tilde\rho \;=\; \alpha\circ\varepsilon_{\mathbb{A}}(\rho)\;=\;\frac{1}{|G|}\sum_{g\in G} V_g\,\rho\,V_g^{\dagger}\;=\;\frac{1}{N}\sum_{r} p_r\sum_{g}\Pi_{r+g}\;=\;\frac{1}{N}\,\id\;=\;\frac{1}{\lambda}\sum_r\Pi_r\, ,
\end{equation}}
{\noindent where we used $\sum_{g}\Pi_{r+g}=\id$ and $\sum_r p_r=1$, the finite-group Haar measure being the uniform weight $1/|G|$. Thus $\alpha\circ\varepsilon_{\mathbb{A}}$ is the canonical (trace-preserving) conditional expectation onto the fixed-point subalgebra; its Watatani index is the order $|G|=N=\lambda$, the classical value for a group-fixed-point inclusion. Then $S_{\cA}(\rho|\tilde\rho)=\log|G|-H(p)$ is simply the distance from $\rho$ to its Haar-averaged, maximally mixed image.}
\medskip 

When $\rho$ is defined through the probability distribution $\bar{p}_r = d_r^2/|G|\, ,  \, \forall\,  r$,  the entropic order parameter vanishes. In this case, the probability distribution $\bar{p}_r$ defines a symmetric phase \cite{Casini:2020}. In this same line, one might consider perturbations around this symmetric phase as density matrices $\rho$ given in terms of 
\begin{align}
            p_r &= \bar{p}_r + \delta p_r\, \quad {\rm s.t}\, \quad \sum_r\, \delta p_r =0\, .
        \end{align}

It is easy to find that for the case $\delta p_r \ll 1$, the entropic order parameter is given by  
         \begin{equation}
             S_{\cA}(\rho|\alpha \circ \varepsilon_{\mathbb{A}}(\rho)) \sim \sum_r\, (\delta p_r)^2\, ,
             \label{eq:rel_pert}
         \end{equation}
in accordance with the positivity of the relative entropy.
\medskip

It is important to note that these patterns may occur due to the presence of non-vanishing one-point functions of non-local global symmetry operators, the only operators that take different expectation values in different phases \cite{Casini:2020, Schafer-Nameki:2025}.  {For $\mathbb{Z}_N$, the relevant non-local operators are the symmetry generators themselves. The group $\mathbb{Z}_N$ is generated by a single unitary $U$ that acts on each charge sector by a phase,}
{\begin{equation}
U=\sum_{r=0}^{N-1}\omega^{\,r}\,\Pi_r\, ,\qquad U\,\Pi_r=\omega^{\,r}\,\Pi_r\, ,\qquad \omega=e^{2\pi i/N}\, ,
\end{equation}}
{\noindent its powers $U^k$ ($k=1,\dots,N-1$) being the nontrivial group elements. Here the index $r\in\{0,\dots,N-1\}$ is the $\mathbb{Z}_N$ charge labelling the sector $\Pi_r$, the index $k$ labels the group element (equivalently, the character of $\mathbb{Z}_N$), and $\omega^{\,kr}$ is the value of the $k$-th character on charge $r$. In a lattice or field-theoretic realization $U$ is an extended (topological) operator rather than a product of strictly local terms and it is the operator conjugate to the charge-shift $V_g$ mentioned above, obeying the shift relation $U\,V_g=\omega^{\,g}\,V_g\,U$. Its expectation values in a state $\rho=\sum_r p_r\,\Pi_r$ are the discrete Fourier components of the charge distribution,}
{\begin{equation}
\langle U^k\rangle_\rho=\mathrm{tr}\,(\rho\,U^k)=\sum_{r=0}^{N-1} p_r\,\omega^{\,kr}\, ,
\end{equation}}
{\noindent and they are exactly what the Haar average erases. On the condensed state $\tilde\rho=\tfrac1N\,\id$ one has $\langle U^k\rangle_{\tilde\rho}=\tfrac1N\sum_r\omega^{\,kr}=0$ for every $k\neq0$. The order parameter measures how far these symmetry-operator expectations sit from zero. Using $S_{\cA}=\log N-H(p)$ together with Parseval's identity}
{\begin{equation}
\sum_{r=0}^{N-1}\Big(p_r-\tfrac1N\Big)^2=\frac1N\sum_{k=1}^{N-1}\big|\langle U^k\rangle_\rho\big|^2\, ,
\end{equation}}
{\noindent one can show that the entropic order parameter grows quadratically in these expectation values near the symmetric point,}
{\begin{equation}
S_{\cA}(\rho|\tilde\rho)\;\simeq\;\frac12\sum_{k=1}^{N-1}\big|\langle U^k\rangle_\rho\big|^2\, .
\label{eq:ZN_quadratic}
\end{equation}}

Said this, it is expected that the behavior of the entropic order parameter behaves quite differently for scenarios where $\delta\, p_r \sim \bar{p}_r$. For example, in the case of the group $\mathbb{Z}_2$ we have
\begin{align*}
           \lbrace{p\rbrace}&=\lbrace 1/2, 1/2 \rbrace \implies S_{\cA}(\lbrace p \rbrace |\lbrace \tilde{p} \rbrace )=0\,   \\ \nonumber
            \lbrace{p\rbrace}&=\lbrace 1, 0\rbrace \implies S_{\cA}(\lbrace p \rbrace |\lbrace \tilde{p} \rbrace )=\log 2\, \\ \nonumber
            \lbrace{p\rbrace}&=\lbrace 0, 1\rbrace\implies S_{\cA}(\lbrace p \rbrace |\lbrace \tilde{p} \rbrace )=\log 2\,
\end{align*}

{This means that if the system starts in state $\lbrace{p\rbrace}=\lbrace 1, 0\rbrace$ or $\lbrace{p\rbrace}=\lbrace 0, 1\rbrace$ post-condensation measurements cannot recover the original state, losing $\log 2$ bits of information—the maximum possible for this condensation process. In addition, the state defined by $\lbrace{p\rbrace}=\lbrace 1/2, 1/2\rbrace$ is invariant, that is to say $\rho = \alpha \circ \varepsilon_{\mathbb{A}}(\rho)$}.

\subsection{The toric code.} 
\label{sec:toric}
{The toric code \cite{Kitaev:1997} is the standard toy model for topological phases. It describes a 2D lattice system with anyonic excitations and 4-fold degenerate ground state manifold, serving as a topological quantum error-correcting code.} In the toric code, the set of anyons is given by $\cA=\lbrace 1, Y, Z, X \rbrace$ with quantum dimensions $d_1= d_Y= d_Z= d_X=1$. The nontrivial fusion rules are those of $\mathbb{Z}_2 \times \mathbb{Z}_2$:
$X \otimes X = Y \otimes Y = Z \otimes Z = 1, X \otimes Y = Z, X \otimes Z = Y, Y \otimes Z = X$.  
In standard physics notation the anyons are labeled as $Y=e$ (electric charge at vertex), $Z=m$ (magnetic flux through a plaquette) and $X=\psi$ (fermionic bound state of $e$ and $m$).
\vspace{0.25 cm}

We consider the pattern of symmetry reduction induced by the algebra $\mathbb{A}=1+Y$. {Physically, this corresponds to allowing electric charges (e-particles) to proliferate freely, screening out their topological distinction from the vacuum}. This leads to a condensed theory $\cT=\lbrace \phi\, ,\mathbb{X}\rbrace \cong {\rm Vec}(\mathbb{Z}_2) $ with quantum dimensions $d_\phi= d_{\mathbb{X}}=1$. 
\medskip

\paragraph*{Determination of the branching coefficients.}
 It is instructive to run the algebraic determination of the branching coefficients explicitly in this example.

 The condensation is specified by $\mathbb A=1+Y$, that is, the vacuum column $n^{\phi}_{1}=n^{\phi}_{Y}=1$ and $n^{\phi}_{Z}=n^{\phi}_{X}=0$. Since only the anyon species $c=1$ and $c=Y$ contribute to the condensation, \eqref{eq:Mprime-from-A} reduces to $M'_{ab}=N^{a}_{b1}+N^{a}_{bY}=\delta_{ab}+\delta_{a,\,b\times Y}$. Fusion with $Y$ acts as the involution $1\leftrightarrow Y$,
$Z\leftrightarrow X$, so that, in the ordering $(1,Y,Z,X)$,
\begin{equation}
  M'=\begin{pmatrix}
  1&1&0&0\\ 1&1&0&0\\ 0&0&1&1\\ 0&0&1&1
  \end{pmatrix}.
\end{equation}
We note that the first column reproduces $n^{\phi}=(1,1,0,0)$, and $M'_{11}=1$.
 $M'$ is block diagonal with two blocks of rank one, so $M'=\Lambda\Lambda^{\!\top}$ requires exactly two columns,
one per block, and
\begin{equation}
    \Lambda=\left(\begin{matrix}1&0\\1&0\\0&1\\0&1\end{matrix}\right)
\end{equation}
up to a relabeling of the columns, which we call $\phi$ and
$\mathbb{X}$. This is the branching $1,Y\to\phi$ and $Z,X\to\mathbb{X}$
used above. All row sums equal one, so $\Lambda\,\vec d_{\cT}=\vec d_{\cA}$ with $d_{\phi}=d_{\mathbb X}=1$ and $\Lambda^{\!\top}\vec d_{\cA}=(2,2)=\lambda\,\vec d_{\cT}$ and $\lambda=\Vert\Lambda\Vert^{2}=2$. The inclusion graph of this algebra inclusion is displayed in Fig.~\ref{fig:toric-graph}. \footnote{This is a bipartite graph where vertices represent anyon species, and edges represent fusion multiplicities, that is, what we call branching coefficients. 
}.

    \begin{figure}
    \centering
    \begin{tikzpicture}
    \draw (-0.75,-3.5) rectangle (3.5,0.5);
    \draw [fill=orange, opacity=0.5] (-0.75,-3.5) rectangle (1.5,0.5);
    \draw [fill=yellow,opacity=0.5] (1.5,-3.5) rectangle (3.5,0.5);
   
    \node(CA) at (0,-4.0) {$\cA$};
    \node(CT) at (3,-4.0) {$\cT$};
    
  \node [circle, draw, fill=red!20](A) at (0,0) {$1$};
  \node [circle, draw, fill=red!20] (B) at (0,-1) {$Y$};
  \node [circle, draw, fill=red!20] (C) at (0,-2) {$X$};
  \node [circle, draw, fill=red!20] (D) at (0,-3) {$Z$};

  \node [circle, draw, fill=blue!20] (E) at (3,-1) {$\phi$};
  \node [circle, draw, fill=blue!20](F) at (3,-2) {$\mathbb{X}$};
 
  \draw[red, very thick, ->] (A) -- (E) node[midway, right] {};
  \draw[red, very thick, ->] (B) -- (E) node[midway, right] {};
  \draw[blue, very thick, ->] (C) -- (F) node[midway, left] {};
  \draw[blue, very thick,->] (D) -- (F) node[midway, above] {};

\end{tikzpicture}        
\caption{Inclusion graph for the symmetry reduction pattern induced by the algebra $\mathbb{A}=1+Y$ in the toric code $\cA \cong$ $\mathbb{Z}_2 \times \mathbb{Z}_2$. This leads to a condensed theory $\cT \cong {\rm Vec}(\mathbb{Z}_2)$}
\label{fig:toric-graph}
\end{figure}

 
With $M'$, we obtain an explicit form for the entropic order parameter
    
    \begin{align}
    \label{eq:toric_S}
             S_{\cA}(\rho | \widetilde\rho)&=\log \lambda -H(p) -\sum_{r \in \lbrace1, Y\rbrace}p_r\, \log p_{\phi} -\sum_{s \in \lbrace X, Z\rbrace}p_s\, \log p_{\mathbb{X}}\\ \nonumber
             &= \log \lambda -H(p) -(p_1+p_Y) \log (p_1+p_Y) -(p_X + p_Z) \log (p_X + p_Z) \, . 
        \end{align}
\medskip

Some examples of initial states $\rho \in \cA$ with their corresponding probabilities $\lbrace{p\rbrace}$ and the value of the entropic order parameter that measures the degree of information loss after condensation are,

\begin{align*}
           \lbrace{p\rbrace}&=\lbrace 1/2, 1/2,0,0 \rbrace \implies S_{\cA}(\lbrace p \rbrace |\lbrace \tilde{p} \rbrace )=0\,   \\ \nonumber
            \lbrace{p\rbrace}&=\lbrace 1/2, 0,0,1/2 \rbrace \implies S_{\cA}(\lbrace p \rbrace |\lbrace \tilde{p} \rbrace )=\log 2\,   \\ \nonumber
            \lbrace{p\rbrace}&=\lbrace 1/3, 1/3,1/3,0 \rbrace +{\rm perms}\implies S_{\cA}(\lbrace p \rbrace |\lbrace \tilde{p} \rbrace )=\frac{1}{3}\log 2\, . \\ \nonumber
\end{align*}

\paragraph{{Lattice realization of the condensation channel and the Entropic Order Parameter.}}
{Now, in order to illustrate the meaning of our construction from a less abstract perspective--the level of abstract charge labels--, we now realize the whole construction --- the charge algebra $\cA$, its projectors $\Pi_a$, and the conditional expectation $\varepsilon_{\mathbb{A}}$ --- explicitly on the physical Hilbert space of the lattice model (Fig.~\ref{fig:toric_lattice}).} {The toric code is built by placing a qubit on every edge $j$ of a square lattice, with vertex and plaquette stabilizers}
{\begin{equation}
A_v = \prod_{j \in v}\sigma^x_j\, , \qquad B_p = \prod_{j \in \partial p}\sigma^z_j\, ,
\end{equation}}
{\noindent whose common $+1$ eigenstate is the ground state. Here $j\in v$ runs over the edges $j$ incident to the vertex $v$ (those having $v$ as an endpoint) and $j\in\partial p$ over the four edges bounding the plaquette $p$.}
\medskip

{An electric charge $e\equiv Y$ is a vertex defect ($A_v=-1$) created at the ends of an open $\sigma^z$ string on the direct lattice; a magnetic flux $m\equiv Z$ is a plaquette defect ($B_p=-1$) created at the ends of an open $\sigma^x$ string on the dual lattice; and $\psi\equiv X = e\times m$. These are the four superselection sectors of $\cA$ realized as genuine lattice excitations.}

{In order to connect with the charge algebra $\cA$, we choose a region $R$. The total anyonic charge it encloses is read off by two commuting closed operators given by,}
{\begin{equation}
W_e(\partial R) = \prod_{j\in\partial R}\sigma^z_j = \prod_{p\in R}B_p\, , \qquad W_m(\partial R) = \prod_{j\in\widetilde{\partial R}}\sigma^x_j = \prod_{v\in R}A_v\, ,
\end{equation}}
{\noindent a Wilson loop on $\partial R$ and a 't~Hooft loop on the dual boundary $\widetilde{\partial R}$ which amounts to the products of the stabilizers enclosed by $\partial R$.} {Being products of Pauli operators, both are Hermitian involutions such that, $W_e^2=W_m^2=\id$, and they commute. Each therefore has eigenvalues $\pm1$, and the two can be diagonalized simultaneously. For any involution $W$, the combination $\tfrac12(\id\pm W)$ projects onto its $\pm1$ eigenspace. Therefore, as $W_e$ and $W_m$ commute, the product of two such projectors,}
{\begin{equation}
\Pi_a \;=\; \tfrac12\big(\id+s^e_a\,W_e\big)\,\tfrac12\big(\id+s^m_a\,W_m\big) \;=\; \tfrac14\big(\id+s^e_a\,W_e\big)\big(\id+s^m_a\,W_m\big)\, ,
\end{equation}}
{\noindent projects onto the simultaneous eigenspace of $W_e$ and $W_m$ with eigenvalues $s^e_a$ and $s^m_a$, respectively. These four joint eigenspaces are exactly the four charge sectors. The Wilson loop detects the enclosed magnetic flux, so $s^e_a=-1$ precisely when $a$ contains $m$, while the 't~Hooft loop detects the enclosed electric charge, so $s^m_a=-1$ when $a$ contains $e$, fixing the signs labelling the four sectors}
{\begin{equation}
(s^e_a,s^m_a)=(+,+),\,(+,-),\,(-,+),\,(-,-)\qquad\text{for}\quad a=1,\,Y,\,Z,\,X,
\end{equation}}
{\noindent in agreement with the charge content of each anyon. The four projectors are orthogonal and resolve the identity, $\sum_a\Pi_a=\id$. In this sense, the abelian algebra $\cA=\bigoplus_a\mathbb{C}\,\Pi_a$ that we have considered abstractly above is, on the lattice, generated by the two loop operators $W_e$ and $W_m$.}
\medskip

\begin{figure}[t]
\centering
\begin{tikzpicture}[scale=0.95, qubit/.style={circle,fill=black,inner sep=1.1pt}]
  \begin{scope}
  \foreach \i in {0,1,2,3}{
    \draw[gray!55] (\i,0)--(\i,3);
    \draw[gray!55] (0,\i)--(3,\i);
  }
  \foreach \y in {0.5,1.5,2.5}{\foreach \x in {0,1,2,3}{\node[qubit] at (\x,\y){};}}
  \foreach \x in {0.5,1.5,2.5}{\foreach \y in {0,1,2,3}{\node[qubit] at (\x,\y){};}}
  \draw[red,very thick] (0,2)--(2,2);
  \draw[red,very thick] (1,1)--(1,3);
  \foreach \p in {(0.5,2),(1.5,2),(1,1.5),(1,2.5)}{\node[red,fill=white,inner sep=0.5pt] at \p {\scriptsize$x$};}
  \fill[red] (1,2) circle (1.6pt);
  \node[red] at (1.35,2.32){$A_v$};
  \draw[blue,very thick] (2,0)--(3,0)--(3,1)--(2,1)--cycle;
  \foreach \p in {(2.5,0),(3,0.5),(2.5,1),(2,0.5)}{\node[blue,fill=white,inner sep=0.5pt] at \p {\scriptsize$z$};}
  \node[blue] at (2.5,0.5){$B_p$};
  \node at (1.5,-0.7){(a)};
  \end{scope}
  \begin{scope}[xshift=5.4cm]
  \foreach \i in {0,1,2,3,4}{
    \draw[gray!55] (\i,0)--(\i,4);
    \draw[gray!55] (0,\i)--(4,\i);
  }
  \fill[orange!18] (1,1) rectangle (3,3);
  \node at (2.65,2.62){\small $R$};
  \draw[blue,very thick] (1,1)--(3,1)--(3,3)--(1,3)--cycle;          
  \node[blue] at (3.45,2){$W_e$};
  \draw[red,very thick,dashed] (1.5,1.5)--(2.5,1.5)--(2.5,2.5)--(1.5,2.5)--cycle; 
  \node[red] at (1.18,1.5){$W_m$};
  \fill[black] (2,2) circle (2.6pt);
  \node at (2.0,2.0)[below right]{\small $e$};
  \draw[blue,very thick,dotted] (2,4)--(2,2);                        
  \node[blue] at (2.0,3.55)[right]{$S_e$};
  \node at (2,-0.7){(b)};
  \end{scope}
\end{tikzpicture}
\caption{Toric code on the square lattice (qubits, black dots, on edges). \textbf{(a)} The vertex operator $A_v=\prod_{j\in v}\sigma^x_j$ (red, $X$-type) and the plaquette operator $B_p=\prod_{j\in\partial p}\sigma^z_j$ (blue, $Z$-type). \textbf{(b)} The anyonic charge enclosed in a region $R$ is read off by two commuting operators: the Wilson loop $W_e=\prod_{j\in\partial R}\sigma^z_j=\prod_{p\in R}B_p$ (blue) on the direct lattice, which detects magnetic flux, and the 't~Hooft loop $W_m=\prod_{j\in\widetilde{\partial R}}\sigma^x_j=\prod_{v\in R}A_v$ (red, dashed) on the dual lattice, which detects electric charge. An open $\sigma^z$ string $S_e$ (blue, dotted) creates and transports an electric charge $e$ into $R$; it commutes with $W_e$ and anticommutes with $W_m$, and averaging over it realizes the condensation channel $\varepsilon_{\mathbb{A}}(O)=\tfrac12\big(O+S_e\,O\,S_e^{\dagger}\big)$.}
\label{fig:toric_lattice}
\end{figure}

{Condensing the anyon $e$ amounts to say that its charge is unobservable in the condensed phase, that is, indistinguishable from the trivial vacuum. The operator that creates and transports an anyon excitation $e$ across $R$ is the open $\sigma^z$ string}
{\begin{equation}
S_e(\gamma) = \prod_{j\in\gamma}\sigma^z_j
\end{equation}}
{\noindent along a path $\gamma$ that enters $R$ and crosses $\partial R$ once. It commutes with $W_e$ and anticommutes with $W_m$, so $S_e W_e S_e^{\dagger}=W_e$ and $S_e W_m S_e^{\dagger}=-W_m$. Therefore, conjugation by $S_e$ leaves $W_e$ invariant but flips the sign of $W_m$, permuting the sectors by fusion with $e$ as}
{\begin{equation}
S_e\, \Pi_a\, S_e^{\dagger} = \Pi_{a\times e}\, , \qquad\text{i.e.}\quad \Pi_1\leftrightarrow\Pi_Y\, ,\ \ \Pi_Z\leftrightarrow\Pi_X\, ,
\end{equation}}
{where $\leftrightarrow$ reads as \emph{``identifies''}.} {\noindent The restriction map $\varepsilon_{\mathbb{A}}$ of our framework is then precisely the $\mathbb{Z}_2$ average that implements these identifications,}
{\begin{equation}
    \varepsilon_{\mathbb{A}}(O) \;=\; \tfrac12\!\left(O + S_e\, O\, S_e^{\dagger}\right),
    \qquad O \in \mathcal{M}_R ,
\end{equation}}
{\noindent that is, a unital, trace-preserving and idempotent conditional expectation onto the $S_e$-invariant subalgebra (recall $S_e^2=\id$) where $\mathcal{M}_R$ refers to the algebra of operators defined on $R$.} {Acting on the loop operators it gives 
\begin{equation}
\varepsilon_{\mathbb{A}}(W_e)=W_e\, \quad {\rm and}\, \quad \varepsilon_{\mathbb{A}}(W_m)=0\, .
\end{equation} 
That is, the channel keeps the $m$-detecting loop and discards the $e$-detecting one. The condensed algebra $\cT$ is thus generated by $W_e$ alone, with projectors}
{\begin{equation}
\Pi_\phi=\tfrac12(\id+W_e)=\Pi_1+\Pi_Y\, , \qquad \Pi_{\mathbb{X}}=\tfrac12(\id-W_e)=\Pi_Z+\Pi_X\, ,
\end{equation}}
{\noindent reproducing the branching $1,Y\to\phi$ and $X,Z\to\mathbb{X}$ obtained above, together with the index $|\mathbb{Z}_2|=2=\lambda$.}
\medskip

{From this, it is straightforward to notice that on a diagonal charge state $\rho=\sum_a p_a\Pi_a$, the channel produces exactly the population transfer obtained above, $p_\phi=p_1+p_Y$ and $p_{\mathbb{X}}=p_X+p_Z$, while the symmetrized state $\tfrac12(\rho+S_e\rho S_e^{\dagger})$ corresponds to the lifted state $\tilde\rho=\alpha\circ\varepsilon_{\mathbb{A}}(\rho)$ entering the relative entropy of Eq.~\eqref{eq:toric_S}. The order parameter $S_{\cA}(\rho|\alpha\circ\varepsilon_{\mathbb{A}}(\rho))$ thereby measures how far the charge distribution is from being invariant under the physical insertion of an $e$ excitation. This is transparent in the examples listed above: the distribution with equal weight on $1$ and $e=Y$ is already $S_e$-invariant, hence fully condensed, and gives $S_{\cA}=0$, whereas weight shared between $1$ and $m=Z$ --- sectors that $S_e$ cannot interconvert --- is maximally sensitive and gives $S_{\cA}=\log 2$.} 
\medskip

{With this, let us now illustrate how the entropic order parameter can be seen as an explicit function of the statistics of non-local operators. Since $S_e$ commutes with $W_e$ and anticommutes with $W_m$, condensation preserves the Wilson loop but erases the 't~Hooft loop, $\langle W_e\rangle_{\tilde\rho}=\langle W_e\rangle_{\rho}$ and $\langle W_m\rangle_{\tilde\rho}=0$. Consider a state $\rho=p_1\Pi_1+p_Y\Pi_Y$ supported in the condensed sector, i.e.\ $p_X=p_Z=0$ and $p_\phi=p_1+p_Y=1$ (so $\langle W_e\rangle_\rho=1$). Specializing the general expression \eqref{eq:toric_S} to this case, the last two terms vanish, $(p_1+p_Y)\log(p_1+p_Y)=\log 1=0$ and $(p_X+p_Z)\log(p_X+p_Z)=0$, while $H(p)$ reduces to the binary entropy} 
\begin{equation}
    H_2(p_1)=-p_1\log p_1-(1-p_1)\log(1-p_1)\, ,
\end{equation}
leaving
{\begin{equation}
S_{\cA}(\rho|\tilde\rho)=\log\lambda-H_2(p_1)=\log 2-H_2(p_1)\, .
\end{equation}}
{\noindent The only nonlocal quantity able to differentiate the condensed charges $1$ and $e$ in this sector is the 't~Hooft loop, with 
\begin{equation}
    \langle W_m\rangle_\rho = \sum_a\, p_a\, s^m_a\, ,
\end{equation}
which, in our case is $\langle W_m\rangle_\rho=p_1-p_Y=2p_1-1$, and hence $p_1=\tfrac12\big(1+\langle W_m\rangle_\rho\big)$. Substituting, we obtain}
{\begin{equation}
S_{\cA}(\rho|\tilde\rho)=\log 2-H_2\left(\tfrac{1}{2}\left(1+\langle W_m\rangle_\rho\right)\right)\, ,
\end{equation}}
{\noindent which runs monotonically from $0$ at $\langle W_m\rangle_\rho=0$ (a state already invariant under the condensation) to the maximal value $\log 2=\log\lambda$ at $\langle W_m\rangle_\rho=\pm1$ (a definite eigenstate of the broken loop). For states close to invariance under the condensation, expanding the binary entropy gives
\begin{equation}
  S(\rho\Vert\tilde\rho)=\tfrac12\,\langle W_{m}\rangle_{\rho}^{2}
  +O\!\left(\langle W_{m}\rangle_{\rho}^{4}\right),
\end{equation}
so that the quantity is quadratic in $\langle W_{m}\rangle_{\rho}$, and vanishes with it. The same structure appeared for $\mathbb{Z}_N$ in \eqref{eq:ZN_quadratic}, where the role of $\langle W_{m}\rangle_{\rho}$ is played by the Fourier components
$\langle U^{k}\rangle_{\rho}$ of the charge distribution. The entropic order parameter is thus a quantitative readout of the expectation value of the non-local operator whose statistics the condensation erases.
}
\medskip

{Finally, the electric--magnetic dual condensation $\mathbb{B}=1+Z$ follows verbatim under $\sigma^z\leftrightarrow\sigma^x$ and direct~$\leftrightarrow$~dual lattice, with the magnetic string $S_m(\tilde\gamma)=\prod_{j\in\tilde\gamma}\sigma^x_j$ (cf.\ Appendix~\ref{AppendixB}).}
\medskip

\subsection{Category {\rm Rep}(S$_3$).}
For this category $\cA=\lbrace 1, X, Y \rbrace$ with quantum dimensions $\lbrace d_1=1, d_X=1, d_Y=2 \rbrace$. The non trivial fusion rules are: 
\begin{equation}
    X\otimes X=1,
    \qquad X\otimes Y=Y,
    \qquad Y\otimes Y=1\oplus X\oplus Y,
    \label{eq:fusion_reps3}
\end{equation} 

\subsubsection*{Condensable Algebra $\mathbb{A}=1+X$}
    This condensable algebra generates a new vacuum in which the only nontrivial anyonic excitation that is condensed is $X$. Here we also find it instructive to run the algebraic determination of the branching coefficients explicitly, since in this example an anyon of $\cA$ splits into two species of $\cT$.
    \medskip
    
    For the ordered sectors $(1,X,Y)$, the vacuum column is
    \begin{equation}
      {\rm n}^{\phi}=(1,1,0)^{\mathsf T}.
    \end{equation}
    Using the fusion rules in Eq. \eqref{eq:fusion_reps3}, one obtains for $M'$ in \eqref{eq:Mprime-from-A}
    \begin{equation}
    M'=
 \begin{pmatrix}
  1&1&0\\
  1&1&0\\
  0&0&2
 \end{pmatrix}.
\end{equation}
The relevant factorization is
\begin{equation}
 M'=\Lambda\Lambda^{\mathsf T},
 \qquad
 \Lambda=
 \begin{pmatrix}
  1&0&0\\
  1&0&0\\
  0&1&1
 \end{pmatrix},
 \label{eq:S3-factorization}
\end{equation}
with columns labeling sector in $\cT$ as $(\phi,t_1,t_2)$. It follows that
\begin{equation}
 1,X\longmapsto\phi,
 \qquad
 Y\longmapsto t_1\oplus t_2\, .
\end{equation}
We note that the dimension $d_Y=2$ is accounted for by $Y$ splitting equally into $t_1$ and $t_2$ s.t $\lbrace d_\phi=1, d_{t_1}=1, d_{t_2}=1 \rbrace$, consistently with \eqref{eq.qtdimAT}. Here, the consistency conditions impose algebraically the splitting of the two-dimensional sector $Y$. It is easy to show then that the pattern of symmetry reduction induced by $\mathbb{A}$ implies $\cT=\lbrace \phi, t_1, t_2 \rbrace \cong {\rm Vec}(\mathbb{Z}_3)$\\

    \begin{figure}
   \centering
    \begin{tikzpicture}
    \draw (-0.75,-2.5) rectangle (3.5,0.5);
    \draw [fill=olive, opacity=0.1] (-0.75,-2.5) rectangle (1.5,0.5);
    \draw [fill=orange, opacity=0.1] (1.5,-2.5) rectangle (3.5,0.5);
   
    \node(CA) at (0,-3) {$\cA \cong {\rm Rep}(S_3)$};
    \node(CT) at (3,-3) {$\cT \cong {\rm Vec}(\mathbb{Z}_3)$};
    
  \node [circle, draw, fill=red!20](A) at (0,0) {$1$};
  \node [circle, draw, fill=red!20] (B) at (0,-1) {$X$};
  \node [circle, draw, fill=red!20] (C) at (0,-2) {$Y$};

  \node [circle, draw, fill=cyan!20] (E) at (3,0) {$\phi$};
  \node [circle, draw, fill=cyan!20](F) at (3,-1) {$t_1$};
  \node [circle, draw, fill=cyan!20](G) at (3,-2) {$t_2$};
 
  \draw[red, very thick, ->] (A) -- (E) node[midway, right] {};
  \draw[red, very thick, ->] (B) -- (E) node[midway, right] {};
  \draw[blue, very thick, ->] (C) -- (F) node[midway, left] {};
  \draw[blue, very thick,->] (C) -- (G) node[midway, above] {};
  
\end{tikzpicture}
\caption{Inclusion graph for the symmetry reduction pattern induced by the algebra $\mathbb{A}=1+X$ in $\cA \cong$ Rep(S$_3$). This leads to a condensed theory $\cT \cong {\rm Vec}(\mathbb{Z}_3)$}
\end{figure}

    In this case the only nontrivial branching coefficients are $n^{\phi}_1=n^{\phi}_X=1$ and $n^{t_{1}}_Y=n^{t_{2}}_Y =1$ and the index of the inclusion is 
    \begin{equation}
        \lambda= \sum_{a\in \cA}\, n^{\phi}_a \, d_a = n^{\phi}_1 d_1 + n^{\phi}_X\, d_X = 2\, .
    \end{equation}
{Equivalently, the index can be derived from the inclusion matrix \eqref{eq:S3-factorization} where $\Lambda^{\!\top}\Lambda = \left(\begin{smallmatrix}2&0&0\\0&1&1\\0&1&1\end{smallmatrix}\right)$, has eigenvalues $\{2,2,0\}$ and $\lambda=\Vert\Lambda\Vert^2=2$. 
    With this
   
        \begin{equation}
             S_{\cA}(\rho |\,  \tilde\rho)=\log \lambda -H(p) -\sum_{r \in \lbrace1, X\rbrace}p_r\, \log (p_1 + p_X) -p_Y\, \log {2\ p_Y}
        \end{equation}
Examples of initial states $\rho \in \cA$  and the value of the entropic order parameter are
\begin{align*}
           \lbrace{p\rbrace}&=\lbrace 1, 0,0\rbrace \implies S_{\cA}(\lbrace p \rbrace |\lbrace \tilde{p} \rbrace )=\log 2\, \\ \nonumber
            \lbrace{p\rbrace}&=\lbrace 1/2, 1/2,0 \rbrace \implies S_{\cA}(\lbrace p \rbrace |\lbrace \tilde{p} \rbrace )=0\, \\ \nonumber
            \lbrace{p\rbrace}&=\lbrace 1/2, 0, 1/2 \rbrace \implies S_{\cA}(\lbrace p \rbrace |\lbrace \tilde{p} \rbrace )=\frac{1}{2}\log 2\, \\ \nonumber
            \lbrace{p\rbrace}&=\lbrace 0, 1/2, 1/2 \rbrace \implies S_{\cA}(\lbrace p \rbrace |\lbrace \tilde{p} \rbrace )=\frac{1}{2}\log 2\, \\ \nonumber
            \lbrace{p\rbrace}&=\lbrace 1/3, 1/3, 1/3 \rbrace \implies S_{\cA}(\lbrace p \rbrace |\lbrace \tilde{p} \rbrace )=0.\\ \nonumber
       \end{align*}   

{We note that, if the system starts in the state $\lbrace{p\rbrace}=\lbrace 1, 0, 0\rbrace$  post-condensation measurements cannot recover the original state by irreversibly losing $\log 2$ bits of information---the maximum possible for this condensation process. Other initial states yield a smaller value of information loss after the symmetrization process and we explicitly find two examples of symmetric states that fulfill $\rho = \alpha \circ \varepsilon_{\mathbb{A}}(\rho)$.}
\medskip

\subsubsection*{Condensable Algebra $\mathbb{A}=1+Y$}
    This condensable algebra generates a new vacuum in which the only nontrivial anyonic excitation that is condensed is $Y$. As in the previous case, it is instructive to determine the branching coefficients algebraically. For the ordered sectors $(1,X,Y)$ the vacuum column is
\begin{equation}
  {\rm n}^{\phi}=(1,0,1)^{\mathsf T},
\end{equation}
and, using the fusion rules \eqref{eq:fusion_reps3},
\eqref{eq:Mprime-from-A} reduces to
$M'_{ab}=\delta_{ab}+N^{a}_{bY}$, giving
\begin{equation}
  M'=\begin{pmatrix} 1&0&1\\ 0&1&1\\ 1&1&2 \end{pmatrix}.
\end{equation}
Again the first column reproduces ${\rm n}^{\phi}$ and $M'_{11}=1$. Since
the third row of $M'$ is the sum of the first two, $M'$ has rank two and
$\Lambda$ has two columns, which we label $(\phi,\mathbb{X})$. The
diagonal entries then fix the factorisation: $M'_{11}=M'_{XX}=1$ forces
the rows of $1$ and $X$ to carry a single unit entry each, $M'_{1X}=0$
forces them into different columns, and $M'_{YY}=2$ together with
$M'_{1Y}=M'_{XY}=1$ forces the row of $Y$ to be $(1,1)$. Hence
\begin{equation}
  M'=\Lambda\Lambda^{\mathsf T},\qquad
  \Lambda=\begin{pmatrix} 1&0\\ 0&1\\ 1&1 \end{pmatrix},
\end{equation}
so that $1\mapsto\phi$, $X\mapsto\mathbb{X}$ and
$Y\mapsto\phi\oplus\mathbb{X}$. Here the two-dimensional sector $Y$
splits with one of its parts entering the new vacuum.  In this case the only nontrivial branching coefficients are $n^{\phi}_1=n^{\phi}_Y=1$ and $n^{\mathbb{X}}_X=n^{\mathbb{X}}_Y =1$. The consistency conditions are met: $\Lambda\,\vec d_{\cT}=\vec d_{\cA}$ with $d_{\phi}=d_{\mathbb{X}}=1$, $\Lambda^{\!\top}\vec d_{\cA}=(3,3) =\lambda\,\vec d_{\cT}$. Now, $\Lambda^{\!\top}\Lambda = \left(\begin{smallmatrix}2&1\\1&2\end{smallmatrix}\right)$ with eigenvalues $\{3,1\}$ and $\lambda = \Vert\Lambda\Vert^2 = 3$. It can be checked that the pattern of symmetry reduction induced by $\mathbb{A}$ implies $\cT=\lbrace \phi, \mathbb{X} \rbrace \cong {\rm Vec}(\mathbb{Z}_2) $.
\medskip

         \begin{figure}
    \centering
    \begin{tikzpicture}
    \draw (-0.75,-2.5) rectangle (3.5,0.5);
    \draw [fill=olive, opacity=0.1] (-0.75,-2.5) rectangle (1.5,0.5);
    \draw [fill=cyan, opacity=0.1] (1.5,-2.5) rectangle (3.5,0.5);
   
    \node(CA) at (0,-3) {$\cA \cong {\rm Rep}(S_3)$};
    \node(CT) at (3,-3) {$\cT \cong {\rm Vec}(\mathbb{Z}_2)$};
    
  \node [circle, draw, fill=red!20](A) at (0,0) {$1$};
  \node [circle, draw, fill=red!20] (B) at (0,-1) {$Y$};
  \node [circle, draw, fill=red!20] (C) at (0,-2) {$X$};

  \node [circle, draw, fill=cyan!20] (E) at (3,-0.5) {$\phi$};
  \node [circle, draw, fill=cyan!20](F) at (3,-1.5) {$\mathbb{X}$};

  \draw[red, very thick, ->] (A) -- (E) node[midway, right] {};
  \draw[red, very thick, ->] (B) -- (E) node[midway, right] {};
  \draw[blue, very thick, ->] (B) -- (F) node[midway, left] {};
  \draw[blue, very thick,->] (C) -- (F) node[midway, above] {};
  
\end{tikzpicture}
\caption{Inclusion graph for the symmetry reduction pattern induced by the algebra $\mathbb{A}=1+Y$ in $\cA \cong$ Rep(S$_3$). This leads to a condensed theory $\cT \cong {\rm Vec}(\mathbb{Z}_2)$}
\end{figure}

    With this
        \begin{equation}
             S_{\cA}(\rho | \tilde\rho)=\log \lambda - H(p) -p_1\, \log \left(p_1+ \frac{p_Y}{2}\right) -p_X\, \log \left(p_X+ \frac{p_Y}{2}\right)  -p_Y\, \log 2, 
        \end{equation}

Examples of initial states $\rho \in \cA$ and the value of the entropic order parameter are 
\begin{align*}
           \lbrace{p\rbrace}&=\lbrace 1, 0,0\rbrace \implies S_{\cA}(\lbrace p \rbrace |\lbrace \tilde{p} \rbrace )=\log 3\, \\ \nonumber
            \lbrace{p\rbrace}&=\lbrace 1/2, 1/2,0 \rbrace  \implies S_{\cA}(\lbrace p \rbrace |\lbrace \tilde{p} \rbrace )=\log 3\, \\ \nonumber
             \lbrace{p\rbrace}&=\lbrace 1/2, 0, 1/2 \rbrace \implies S_{\cA}(\lbrace p \rbrace |\lbrace \tilde{p} \rbrace )=\frac{1}{2}\log 3/2\, \\ \nonumber
             \lbrace{p\rbrace}&=\lbrace 0, 1/2, 1/2 \rbrace \implies S_{\cA}(\lbrace p \rbrace |\lbrace \tilde{p} \rbrace )=\frac{1}{2}\log 3/2\, \\ \nonumber
              \lbrace{p\rbrace}&=\lbrace 1/3, 1/3, 1/3 \rbrace \implies S_{\cA}(\lbrace p \rbrace |\lbrace \tilde{p} \rbrace )=\frac{1}{3}\log 2. \\ \nonumber
       \end{align*}
{We note that, if the system starts in the state $\lbrace{p\rbrace}=\lbrace 1, 0, 0\rbrace$  post-condensation measurements cannot recover the original state by irreversibly losing $\log 3$ bits of information---the maximum possible for this condensation process. Other initial states yield a smaller value of information loss after the symmetrization process.}
\medskip

  \subsubsection*{Condensable Algebra $\mathbb{A}=1+X +2Y$}  
    This is a Lagrangian algebra. As in the case of the group, the pattern of symmetry reduction induced by $\mathbb{A}$ yields a new vacuum $\phi$ in which all the nontrivial anyonic charges $a$ have condensed, and, this implies that $\cT=\lbrace \phi\rbrace \cong {\rm Vec}(\mathbb{C}) $ with $\lbrace d_\phi=1 \rbrace$\\
    
     \begin{figure}[h]
    \centering
    \begin{tikzpicture}
    \draw (-0.75,-2.5) rectangle (3.5,0.5);
    \draw [fill=olive, opacity = 0.1] (-0.75,-2.5) rectangle (1.5,0.5);
    \draw [fill=pink,  opacity = 0.05] (1.5,-2.5) rectangle (3.5,0.5);
   
    \node(CA) at (0,-3) {$\cA \cong {\rm Rep}(S_3)$};
    \node(CT) at (3,-3) {$\cT \cong {\rm Vec}(\mathbb{C})$};
    
  \node [circle, draw, fill=red!20](A) at (0,0) {$1$};
  \node [circle, draw, fill=red!20] (B) at (0,-1) {$X$};
  \node [circle, draw, fill=red!20] (C) at (0,-2) {$Y$};

  \node [circle, draw, fill=cyan!20] (E) at (3,-1) {$\phi$};
  
  \draw[red, very thick, ->] (A) -- (E) node[midway, right] {};
  \draw[red, very thick, ->] (B) -- (E) node[midway, right] {};
  \draw[red, very thick, ->] (C) --(E) node[midway, left] {};
  
\end{tikzpicture}
\caption{Inclusion graph for the symmetry reduction pattern induced by the algebra $\mathbb{A}=1+X + 2Y$ in $\cA \cong$ Rep(S$_3$). This leads to a trivial condensed theory $\cT \cong {\rm Vec}(\mathbb{C})$}
\end{figure}
    
    In this case the only nontrivial branching coefficients are $n^{\phi}_1=n^{\phi}_X=1$ and $n^{\phi}_Y=2$ and the index of the inclusion is 
    \begin{equation}
        \lambda= \sum_{a\in \cA}\, n^{\phi}_a \, d_a = n^{\phi}_1 d_1 + n^{\phi}_X\, d_X + n^{\phi}_Y\, d_Y = 6\, .
    \end{equation}
{Being Lagrangian ($\cT=\{\phi\}$), the inclusion matrix \eqref{eq:inclusionmatrix} is the single column $\Lambda=(1,1,2)^{\!\top}$ (rows $1,X,Y$), so $\Lambda^{\!\top}\Lambda = 1+1+4 = 6 = \lambda$ and $M'=\Lambda\Lambda^{\!\top}=\left(\begin{smallmatrix}1&1&2\\1&1&2\\2&2&4\end{smallmatrix}\right)$; the multiplicity $n^{\phi}_Y=2$ enters directly as the matrix entry, and $\Vert\Lambda\Vert^2 = \mathcal{D}_{\cA}^2 = 6$.}
    With this

        \begin{equation}
             S_{\cA}(\rho | \tilde\rho)=\log \lambda - H(p) -2\, p_Y\, \log 2\, 
        \end{equation}
 Some examples of initial states $\rho \in \cA$ and the value of the entropic order parameter are
\begin{align*}
           \lbrace{p\rbrace}&=\lbrace 1, 0,0\rbrace \implies S_{\cA}(\lbrace p \rbrace |\lbrace \tilde{p} \rbrace )=\log 6\, \\ \nonumber
             \lbrace{p\rbrace}&=\lbrace 1/2, 1/2,0 \rbrace  \implies S_{\cA}(\lbrace p \rbrace |\lbrace \tilde{p} \rbrace )=\log 3\, \\ \nonumber
              \lbrace{p\rbrace}&=\lbrace 1/2, 0, 1/2 \rbrace \implies S_{\cA}(\lbrace p \rbrace |\lbrace \tilde{p} \rbrace )=\log 3/2\, \\ \nonumber
              \lbrace{p\rbrace}&=\lbrace 0, 1/2, 1/2 \rbrace \implies S_{\cA}(\lbrace p \rbrace |\lbrace \tilde{p} \rbrace )=\log 3/2\, \\ \nonumber
               \lbrace{p\rbrace}&=\lbrace 1/3, 1/3, 1/3 \rbrace \implies S_{\cA}(\lbrace p \rbrace |\lbrace \tilde{p} \rbrace )=\frac{1}{3}\log 2. \\ \nonumber
       \end{align*}

{That is, if the system starts in the state $\lbrace{p\rbrace}=\lbrace 1, 0, 0\rbrace$  post-condensation measurements cannot recover the original state by irreversibly losing $\log 6$ bits of information---the maximum possible for this condensation process. Other initial states yield a smaller value of information loss after the symmetrization process.}
    \medskip

\begin{figure}[h]\centering
\begin{tikzpicture}
\draw [cyan,  fill=cyan, opacity = 0.1]   
(-1.5, -2) -- (7, -2) -- (7, -5.5) -- (-1.5,-5.5) --(-1.5,-2) ;
\draw [orange,  fill=orange, opacity = 0.1]   
(7, 1.5) -- (7, -5.5) -- (2.5,-5.5) --(2.5,1.5) -- (7,1.5);
\node[fill= white, draw,rectangle,thick,align=center](Trivial) at 
(0.5,0) 
{{Trivial Phase} \\  
\begin{tikzpicture}
\node at (2,0) {\textcolor{red}{Vec($\mathbb{C}$)}};
\end{tikzpicture} };
\node[fill= white,draw,rectangle,thick,align=center](RepS3Z2) at 
(5,0) 
{{${\rm Rep} (S_3)/\mathbb{Z}_2$ Phase} \\ 
\centering
\begin{tikzpicture}
\node at (2,0) {\textcolor{red}{Vec($\mathbb{Z}_3$)}};
\end{tikzpicture}
};
\node[fill= white, draw,rectangle,thick,align=center](Z2SSB) at (0.5,-4) {{$\mathbb{Z}_2$ SSB Phase} \\  
\begin{tikzpicture}[baseline]
\node at (0.7,0) {\textcolor{red}{Vec($\mathbb{Z}_2$)}};
\begin{scope}[shift={(0.3, 0)}]
\end{scope}
\begin{scope}[shift={(2.7, 0)}]
\end{scope}
\end{tikzpicture}
};
\node[fill= white,draw,rectangle,thick,align=center](RepS3) at (5,-4) {{${\rm Rep}
(S_3)$ Phase} \\ 
\begin{tikzpicture}[baseline]
\node at (0,0) {\textcolor{red}{Rep($S_3$)}};
\begin{scope}[shift={(2, 0)}]
\end{scope}
\end{tikzpicture}
};
\draw[teal, thick,stealth-stealth] (Trivial) to (RepS3Z2);
\draw[violet, thick,stealth-stealth] (RepS3) to (RepS3Z2);
\draw[blue, thick, stealth-stealth] (Trivial) to (Z2SSB);
\draw[purple, thick,stealth-stealth] (Z2SSB) to (RepS3);
\draw[red, thick,stealth-stealth] (Trivial) to (RepS3);
   \node[right] at (5,-2.25) {\small \textcolor{violet}{$\mathbb{A}$=1 + X}};
  \node at (2.4,-4.5) {\small \textcolor{purple}{$\mathbb{A}$=1+Y}}; 
   \node at (2.3,-2.7) {\small \textcolor{red}{$\mathbb{A}$=1+X + 2Y}}; 
\begin{scope}[shift={(1,-0.1)}]
\end{scope}
\begin{scope}[shift={(2.1,-5.3)}]
\end{scope}
\end{tikzpicture}
\caption{Patterns of ${\rm Rep}(S_3)$ symmetry reduction between different phases are indicated by arrows with their corresponding condensation algebras. The symmetries of each phase are given in red. Symmetry reduction from Vec($\mathbb{Z}_N$) to Vec($\mathbb{C}$) are implemented by the group-like condensable algebra discussed in the main text. 
}
\label{fig:reps3_ssb}
\end{figure}
{The different patterns of symmetry reduction in Rep(S$_3$) are summarized in Fig. \ref{fig:reps3_ssb}.} {We note that in $(1+1)$-d, Rep(S$_3$) gapped and gapless phases were found using continuum methods in \cite{Bhardwaj:2023idu, Bhardwaj:2023bbf, Bhardwaj:2024qrf}. In \cite{Bhardwaj:2024}, the authors provided lattice models as concrete realizations of these Rep(S$_3$) phases that were also discussed in \cite{Eck:2023gic}.}

\section{Conclusions}
In this work, we have introduced an operator–algebraic and information–theoretic framework to study {patterns of (generalized) symmetry reduction in theories with generalized symmetries} in $(2+1)$-dimensions. The central idea is to model the restriction and lifting operations associated with a condensable Frobenius algebra by means of conditional expectations and coarse–graining maps. These maps admit explicit Kraus representations determined by the branching coefficients of the condensation, making the construction transparent and computationally tractable.
\medskip

From this setting, we extracted an information–theoretic order parameter: the quantum relative entropy between a state and its lifted image. This quantity measures the loss of distinguishability induced by condensation, and is bounded universally by the logarithm of {the index of conditional expectation}, which coincides with the quantum dimension of the condensate. In this way, the amount of information erased in the restriction process is directly tied to a fundamental algebraic invariant. Where the
construction reaches the lattice operator algebra, it is a readout of the
expectation value of the extended operator the condensation erases. This gives a precise bridge between algebraic structures underlying anyon condensation and entropic measures familiar from quantum information theory.
\medskip

We illustrate the general construction through a variety of examples, including abelian $\mathbb{Z}_N$ models, the toric code, and Rep$(S_3)$. In each case, the formalism reproduces the expected condensed phases, while providing closed–form entropic expressions and saturation of the universal bound in special regimes. We also clarified how dualities correspond to equivalent condensations at the categorical level, showing that our formalism is compatible with known equivalence relations in topological order.
\medskip

{We emphasize that {the finite-dimensional abelian algebras $\cA$ and $\cT$ used throughout are the charge-sector truncation of the genuine subfactor that the condensation defines on the local algebras of the 2D system. The invariant we compute is the Watatani index of the associated conditional expectation.  Carrying out the full type II$_1$/III treatment on the local (cone) algebras of an explicit lattice model, in the spirit of \cite{Naaijkens:2018, Fiedler:2016}, is the natural next step and would promote the lattice channel of Section~\ref{sec:toric} to a genuine subfactor conditional expectation.}}
\medskip

In summary, the combination of {an algebraic approach to anyon condensation} and quantum information tools provides a powerful and versatile language for understanding {generalized symmetry reduction}, yielding both conceptual insights and concrete computational tools.

\section*{Acknowledgements}
We thank Marina Huerta, Horacio Casini, and Javier M. Mag\'an for the stimulating discussions and their kind invitation to participate in the workshop \emph{“Anomalies, Topology and Quantum Information in Field Theory and Condensed Matter Physics”} 
(ICTP-SAIFR, São Paulo, Brazil, June 16–27, 2025). 


\paragraph{Funding information}
J.M.-V. thanks the financial support of Spanish MINECO grant PID2024-155685NB-C22 and Fundaci\'on S\'eneca de la Regi\'on de Murcia FSRM/10.13039/100007801 (22581/PI/24). G.S. and H-C.Z. acknowledge financial support from the Spanish MINECO grant PID2021-127726NB-I00, the CSIC Research Platform on Quantum Technologies PTI-001 and the QUANTUM ENIA project Quantum Spain funded through the RTRP-Next Generation program under the framework of the Digital Spain 2026 Agenda. H.-C.Z. acknowledges support by an appointment to the Young Scientist Training (YST) Program at Asia Pacific Center for Theoretical Physics (APCTP) through the Science and Technology Promotion Fund and Lottery Fund of the Korean Government, and support by the Korean Local Governments - Gyeongsangbuk-do Province and Pohang City.

\begin{appendix}
\numberwithin{equation}{section}

\section{Idempotence and Bimodularity of $\varepsilon_{\mathbb{A}}$}
\label{AppendixA}
In this appendix we prove two central properties of the conditional expectation $\cE_{\mathbb{A}}$ defining the inclusion $\cT \subset \cA$: the idempotence, which amounts to say that $\cE_{\mathbb{A}}$ is a projector from $\cA$ to $\cT$; and the bimodule property.
\paragraph{Idempotence.}
We prove that the operator $\varepsilon_{\mathbb{A}}$ is idempotent by computing $\varepsilon_{\mathbb{A}}\left(\tilde \rho\right)\in \cT $ with $\tilde \rho = \alpha(\rho_{\mathbb{A}})$. Graphically this amounts to

\begin{center}
\begin{tikzcd}
\rho  \arrow{r}{\varepsilon_{\mathbb{A}}}  
& 
\rho_{\mathbb{A}}  \arrow{r}{\alpha} 
&
\tilde \rho \arrow{r}{\varepsilon_{\mathbb{A}}}  
&
\varepsilon_{\mathbb{A}}\left(\tilde \rho\right) 
\end{tikzcd}
\end{center}

and then checking if $\varepsilon_{\mathbb{A}}\left(\tilde \rho\right)$ equals $\varepsilon_{\mathbb{A}}\left(\rho\right) $. Note that the second step is necessary as $\varepsilon_{\mathbb{A}}$ acts on operators expressed in terms of $\cA$ elements. Then, as stated in the main text,
\begin{align}
     \varepsilon_{\mathbb{A}}(\rho)&=\sum_{t\in \cT}\, p_t\, \Pi_t\, , \quad  p_t= \sum_{a\in \cA}\, n^t_a\left(\frac{d_t}{d_a}\right)\, p_a\, ,
\end{align}
and 
\begin{align}
     \tilde \rho& =\sum_{a\in \cA}\, \widetilde{p}_a\, \Pi_a \, ,\quad \widetilde{p}_a=\frac{1}{\lambda} \sum_{t\in \cT}\, n^t_a\left(\frac{d_a}{d_t}\right)\, p_t\, .
\end{align}

Therefore, we find that
\begin{align}
     \varepsilon_{\mathbb{A}}\left(\tilde \rho\right)&=\sum_{t\in \cT}\, \tilde p_t\, \Pi_t\, , \quad  \tilde p_t= \sum_{a\in \cA}\, n^t_a\left(\frac{d_t}{d_a}\right)\, \tilde p_a\, .
\end{align}
   
The equality implying idempotence holds if $p_t = \tilde p_t,\, \forall\, t \in \cT$, which immediately follows from Eqs. \eqref{eq.qtdimAT} and \eqref{eq.qtdimTA}.
\medskip

\paragraph{Bimodule property.} This refers to
\[\cE_{\mathbb{A}}( {\tilde p}\,  {m}\,  {\tilde q}) = {p}\,  \cE_{\mathbb{A}}({m})\,  { q}\, \quad  \forall\,   {p}, { q} \in \mathcal{T}, {m} \in \mathcal{A} ,\]

where the tildes refer to the lifted versions of operators in $\cT$. To prove this property, we first focus on the first term where
\begin{align}
 {\tilde p} &= \sum_{b \in \cA}\,  {\tilde p}_b\, \Pi_b\, ,\quad {\tilde q} = \sum_{c \in \cA}\,  {\tilde q}_c\, \Pi_c\, , \quad 
 {m} = \sum_{d \in \cA}\,  {m}_d\, \Pi_d\, .
\end{align}

Now, we note that for $a \in \cA$,
\begin{align}
    \mathbb{K}_a\, {\tilde p} &= \sum_{t\in \cT}\, \sqrt{\lambda^t_a}\, \Delta_{t,a} \left(\sum_{b \in \cA}\,  {\tilde p}_b\, \Pi_b\right) = \sum_{t\in \cT}\, \sum_{b \in \cA}\, \sqrt{\lambda^t_a}\, {\tilde p}_b\, \Delta_{t,a} \, \Pi_b\\ \nonumber
    &=\sum_{t\in \cT}\,  \sqrt{\lambda^t_a}\, {\tilde p}_a\, \Delta_{t,a} \, ,
\end{align}

where $\lambda^t_a=n^t_a\, \left(\frac{d_t}{d_a}\right)$. Similarly,
\begin{align}
    {\tilde q}\, \mathbb{K}^{\dagger}_a = \sum_{t\in \cT}\,  \sqrt{\lambda^t_a}\, {\tilde q}_a\, \Delta^{\dagger}_{t,a} \, .
\end{align}

With this,
\begin{align}
    \cE_{\mathbb{A}}({\tilde p}\, {m}\, {\tilde q})& =\sum_{a \in \cA}\, \mathbb{K}_a\, ({\tilde p}\, {m}\, {\tilde q})\, \mathbb{K}^{\dagger}_a =\sum_{t \in \cT}\, \sum_{a \in \cA}\, \lambda^t_a\, \left({\tilde p}_a {\tilde q}_a\,  {m}_a\right)\, \Delta_{t,a}\Delta^{\dagger}_{t,a} \\ \nonumber
    & = \sum_{t \in \cT}\, \left(\sum_{a \in \cA}\, \lambda^t_a\, \left({\tilde p}_a {\tilde q}_a\,  {m}_a\right)\right)\, \Pi_t = \sum_{t \in \cT}\, c_t\, \Pi_t\, ,
\end{align}
with
\begin{align}\label{eq:bimod_coeff}
    c_t=\sum_{a \in \cA}\, \lambda^t_a\, \left({\tilde p}_a {\tilde q}_a\,  {m}_a\right)\, .
\end{align}

Now we focus on the second term where
\begin{align}
 {p} &= \sum_{r \in \cT}\,  {p}_r\, \Pi_r\, ,\quad  {q} = \sum_{s \in \cT}\, { q}_s\, \Pi_s\, , \quad 
\cE_{\mathbb{A}} ({m}) = \sum_{t \in \cT}\,  {m}_t\, \Pi_t\, ,
\end{align}
in such a way that 
\begin{align}
    {p}\,  \cE_{\mathbb{A}}({m})\,  { q} &= \sum_{t,r,s \in \cT}\,  \left({p}_r\,  { q}_s\,  {m}_t\right)\, \delta_{rs}\delta_{st}\, \Pi_t\\ \nonumber 
    &=\sum_{t \in \cT}\, \left({p}_t\,  { q}_t\,  {m}_t\right)\, \Pi_t =\sum_{t \in \cT}\, c_t\, \Pi_t\, ,
\end{align}
where the last equality uses Eqs. \eqref{eq:ptcond} and \eqref{eq:bimod_coeff}, thus proving the bimodule property of $\cE_{\mathbb{A}}$.

\section{Dualities and equivalent condensations}
\label{AppendixB}
Two condensable algebras are Morita equivalent when their categories of local modules --- the condensed phases $\mathcal{U}$ --- are equivalent as modular tensor categories \cite{Mueger:2001}. Physically, the two condensation-induced ``vacuum redefinitions'' then yield the same long-range phase. We show that, already at the level of the restriction step $\cA\to\cT$, an equivalence between two condensations $\mathbb{A}$ and $\mathbb{B}$ is mediated by a duality of the parent theory --- a sector-permuting symmetry of $\cA$ implemented by a unitary
\begin{equation}
    \mathrm{Q}=\sum_{a,b \in \cA}\, q_{ab}\, X_{ab}\, ,\qquad X_{ab}\, \Pi_b\, X_{ab}^{\dagger}= \Pi_a\, .
\end{equation}
Writing the two condensations through their branching coefficients $(n^t_a)_{\mathbb{A}}$, $(n^t_a)_{\mathbb{B}}$ and the associated Kraus operators $\mathbb{K}^{\mathbb{A}}_a=\sum_{t}\lambda^{\mathbb{A}}_{ta}\Delta_{ta}$ (and similarly for $\mathbb{B}$), with $\lambda^{\mathbb{A}}_{ta}=\sqrt{(n^t_a)_{\mathbb{A}}\,d_t/d_a}$, the duality relates them by
\begin{equation}
    \varepsilon_{\mathbb{A}}(\mathrm{Q}\, \rho\, \mathrm{Q}^{\dagger}) = \varepsilon_{\mathbb{B}}(\rho)\, ,\qquad \lambda_{ta}^{\mathbb{B}} = \sum_{b \in \cA}\, q_{ab}\, \lambda_{tb}^{\mathbb{A}}\, .
\end{equation}

The toric code furnishes the simplest example. The condensation $\mathbb{A}=1+Y$ has nonzero branching $n^{\phi}_1=n^{\phi}_Y=1$ and $n^{\mathbb{X}}_X=n^{\mathbb{X}}_Z=1$ (the inclusion graph of Section~\ref{sec:toric}), giving $\cT=\{\phi,\mathbb{X}\}\cong\mathrm{Vec}(\mathbb{Z}_2)$ and
\begin{equation}
    \varepsilon_{\mathbb{A}}(\rho)= p_{\phi}\, \Pi_{\phi} + p_{\mathbb{X}}\, \Pi_{\mathbb{X}}\, ,\qquad p_\phi = p_1 + p_Y\, ,\quad p_{\mathbb{X}} = p_X + p_Z\, .
\end{equation}
The dual condensation $\mathbb{B}=1+Z$ is obtained by relabeling $Y\leftrightarrow Z$, giving $p_\phi=p_1+p_Z$ and $p_{\mathbb{X}}=p_X+p_Y$.

{In the notation of \eqref{eq:inclusionmatrix}, the dual inclusion matrix is $\Lambda_{\mathbb{B}}=\left(\begin{smallmatrix}1&0\\0&1\\0&1\\1&0\end{smallmatrix}\right)$ (rows $1,Y,X,Z$; columns $\phi,\mathbb{X}$), obtained from the direct one $\Lambda_{\mathbb{A}}=\left(\begin{smallmatrix}1&0\\1&0\\0&1\\0&1\end{smallmatrix}\right)$ by the row permutation $Y\leftrightarrow Z$. Since the norm is invariant under such a permutation, $\Vert\Lambda_{\mathbb{B}}\Vert^2 = 2 = \Vert\Lambda_{\mathbb{A}}\Vert^2$ and the dual condensation carries the same index.} This permutation is implemented by
\begin{equation}
{\rm Q}= \left(\begin{array}{cc} \mathbb{I}_{2}& 0  \\ 0 & X \end{array}\right)\,,\qquad X= \left(\begin{array}{cc} 0 & 1 \\ 1 & 0 \end{array}\right)\,,
\end{equation}
from which one checks directly that $\varepsilon_{\mathbb{A}}(\mathrm{Q}\, \rho\, \mathrm{Q}^{\dagger}) = \varepsilon_{\mathbb{B}}(\rho)$.

\end{appendix}

\bibliography{info_loss_bib}

@article{Gaiotto:2014,
    author = "Gaiotto, Davide and Kapustin, Anton and Seiberg, Nathan and Willett, Brian",
    title = "{Generalized Global Symmetries}",
    eprint = "1412.5148",
    archivePrefix = "arXiv",
    primaryClass = "hep-th",
    doi = "10.1007/JHEP02(2015)172",
    journal = "JHEP",
    volume = "02",
    pages = "172",
    year = "2015"
}

@article{McGreevy:2022,
    author = "McGreevy, John",
    title = "{Generalized Symmetries in Condensed Matter}",
    eprint = "2204.03045",
    archivePrefix = "arXiv",
    primaryClass = "cond-mat.str-el",
    doi = "10.1146/annurev-conmatphys-040721-021029",
    journal = "Ann. Rev. Condensed Matter Phys.",
    volume = "14",
    pages = "57--82",
    year = "2023"
}

@article{Bhardwaj:2023,
    author = "Bhardwaj, Lakshya and Bottini, Lea E. and Pajer, Daniel and Schafer-Nameki, Sakura",
    title = "{Categorical Landau Paradigm for Gapped Phases}",
    eprint = "2310.03786",
    archivePrefix = "arXiv",
    primaryClass = "cond-mat.str-el",
    doi = "10.1103/PhysRevLett.133.161601",
    journal = "Phys. Rev. Lett.",
    volume = "133",
    number = "16",
    pages = "161601",
    year = "2024"
}

@article{Perez-Lona:2023djo,
    author = "Perez-Lona, A. and Robbins, D. and Sharpe, E. and Vandermeulen, T. and Yu, X.",
    title = "{Notes on gauging noninvertible symmetries. Part I. Multiplicity-free cases}",
    eprint = "2311.16230",
    archivePrefix = "arXiv",
    primaryClass = "hep-th",
    doi = "10.1007/JHEP02(2024)154",
    journal = "JHEP",
    volume = "02",
    pages = "154",
    year = "2024"
}

@article{Diatlyk:2023fwf,
    author = "Diatlyk, Oleksandr and Luo, Conghuan and Wang, Yifan and Weller, Quinten",
    title = "{Gauging non-invertible symmetries: topological interfaces and generalized orbifold groupoid in 2d QFT}",
    eprint = "2311.17044",
    archivePrefix = "arXiv",
    primaryClass = "hep-th",
    doi = "10.1007/JHEP03(2024)127",
    journal = "JHEP",
    volume = "03",
    pages = "127",
    year = "2024"
}

@article{Bhardwaj:2023idu,
    author = {Bhardwaj, Lakshya and Bottini, Lea E. and Pajer, Daniel and Sch{\"a}fer-Nameki, Sakura},
    title = "{Gapped phases with non-invertible symmetries: (1+1)d}",
    eprint = "2310.03784",
    archivePrefix = "arXiv",
    primaryClass = "hep-th",
    doi = "10.21468/SciPostPhys.18.1.032",
    journal = "SciPost Phys.",
    volume = "18",
    number = "1",
    pages = "032",
    year = "2025"
}

@article{Aksoy:2025,
    author = {Aksoy, {\"O}mer M. and Wen, Xiao-Gang},
    title = "{Phases with non-invertible symmetries in 1+1D -- symmetry protected topological orders as duality automorphisms}",
    eprint = "2503.21764",
    archivePrefix = "arXiv",
    primaryClass = "cond-mat.str-el",
    month = "3",
    year = "2025"
}

@article{Bhardwaj:2024qiv,
    author = "Bhardwaj, Lakshya and Pajer, Daniel and Schafer-Nameki, Sakura and Tiwari, Apoorv and Warman, Alison and Wu, Jingxiang",
    title = "{Gapped phases in (2+1)d with non-invertible symmetries: Part I}",
    eprint = "2408.05266",
    archivePrefix = "arXiv",
    primaryClass = "hep-th",
    doi = "10.21468/SciPostPhys.19.2.056",
    journal = "SciPost Phys.",
    volume = "19",
    number = "2",
    pages = "056",
    year = "2025"
}

@article{Bhardwaj:2025piv,
    author = "Bhardwaj, Lakshya and Schafer-Nameki, Sakura and Tiwari, Apoorv and Warman, Alison",
    title = "{Gapped Phases in (2+1)d with Non-Invertible Symmetries: Part II}",
    eprint = "2502.20440",
    archivePrefix = "arXiv",
    primaryClass = "hep-th",
    month = "2",
    year = "2025"
}

@article{Inamura:2025cum,
    author = "Inamura, Kansei and Huang, Sheng-Jie and Tiwari, Apoorv and Schafer-Nameki, Sakura",
    title = "{(2+1)d lattice models and tensor networks for gapped phases with categorical symmetry}",
    eprint = "2506.09177",
    archivePrefix = "arXiv",
    primaryClass = "cond-mat.str-el",
    doi = "10.21468/SciPostPhys.20.2.043",
    journal = "SciPost Phys.",
    volume = "20",
    number = "2",
    pages = "043",
    year = "2026"
}

@article{Ji:2019jhk,
    author = "Ji, Wenjie and Wen, Xiao-Gang",
    title = "{Categorical symmetry and noninvertible anomaly in symmetry-breaking and topological phase transitions}",
    eprint = "1912.13492",
    archivePrefix = "arXiv",
    primaryClass = "cond-mat.str-el",
    doi = "10.1103/PhysRevResearch.2.033417",
    journal = "Phys. Rev. Res.",
    volume = "2",
    number = "3",
    pages = "033417",
    year = "2020"
}

@article{Chatterjee:2022tyg,
    author = "Chatterjee, Arkya and Wen, Xiao-Gang",
    title = "{Holographic theory for continuous phase transitions: Emergence and symmetry protection of gaplessness}",
    eprint = "2205.06244",
    archivePrefix = "arXiv",
    primaryClass = "cond-mat.str-el",
    doi = "10.1103/PhysRevB.108.075105",
    journal = "Phys. Rev. B",
    volume = "108",
    number = "7",
    pages = "075105",
    year = "2023"
}

@article{Bhardwaj:2023bbf,
    author = "Bhardwaj, Lakshya and Bottini, Lea E. and Pajer, Daniel and Schafer-Nameki, Sakura",
    title = "{The club sandwich: Gapless phases and phase transitions with non-invertible symmetries}",
    eprint = "2312.17322",
    archivePrefix = "arXiv",
    primaryClass = "hep-th",
    doi = "10.21468/SciPostPhys.18.5.156",
    journal = "SciPost Phys.",
    volume = "18",
    number = "5",
    pages = "156",
    year = "2025"
}

@article{Wen:2023otf,
    author = "Wen, Rui and Potter, Andrew C.",
    title = "{Classification of 1+1D gapless symmetry protected phases via topological holography}",
    eprint = "2311.00050",
    archivePrefix = "arXiv",
    primaryClass = "cond-mat.str-el",
    doi = "10.1103/PhysRevB.111.115161",
    journal = "Phys. Rev. B",
    volume = "111",
    number = "11",
    pages = "115161",
    year = "2025"
}

@article{Bhardwaj:2024qrf,
    author = "Bhardwaj, Lakshya and Pajer, Daniel and Schafer-Nameki, Sakura and Warman, Alison",
    title = "{Hasse diagrams for gapless SPT and SSB phases with non-invertible symmetries}",
    eprint = "2403.00905",
    archivePrefix = "arXiv",
    primaryClass = "cond-mat.str-el",
    doi = "10.21468/SciPostPhys.19.4.113",
    journal = "SciPost Phys.",
    volume = "19",
    number = "4",
    pages = "113",
    year = "2025"
}

@article{Bhardwaj:2025jtf,
    author = "Bhardwaj, Lakshya and Gai, Yuhan and Huang, Sheng-Jie and Inamura, Kansei and Schafer-Nameki, Sakura and Tiwari, Apoorv and Warman, Alison",
    title = "{Gapless Phases in (2+1)d with Non-Invertible Symmetries}",
    eprint = "2503.12699",
    archivePrefix = "arXiv",
    primaryClass = "cond-mat.str-el",
    month = "3",
    year = "2025"
}

@article{Wen:2025thg,
    author = "Wen, Rui",
    title = "{Topological Holography for 2+1-D Gapped and Gapless Phases with Generalized Symmetries}",
    eprint = "2503.13685",
    archivePrefix = "arXiv",
    primaryClass = "hep-th",
    month = "3",
    year = "2025"
}

@article{Burnell:2018,
   title={Anyon Condensation and Its Applications},
   volume={9},
   ISSN={1947-5462},
   url={http://dx.doi.org/10.1146/annurev-conmatphys-033117-054154},
   DOI={10.1146/annurev-conmatphys-033117-054154},
   number={1},
   journal={Annual Review of Condensed Matter Physics},
   publisher={Annual Reviews},
   author={Burnell, F.J.},
   year={2018},
   month=Mar, pages={307–327} }

@article{Bischoff:2018juy,
   title={Spontaneous symmetry breaking from anyon condensation},
   volume={2019},
   ISSN={1029-8479},
   url={http://dx.doi.org/10.1007/JHEP02(2019)062},
   DOI={10.1007/jhep02(2019)062},
   number={2},
   journal={Journal of High Energy Physics},
   publisher={Springer Science and Business Media LLC},
   author={Bischoff, Marcel and Jones, Corey and Lu, Yuan-Ming and Penneys, David},
   year={2019},
   month=Feb }

@article{Casini:2022,
    author = "Casini, Horacio and Huerta, Marina",
    title = "{Lectures on entanglement in quantum field theory}",
    eprint = "2201.13310",
    archivePrefix = "arXiv",
    primaryClass = "hep-th",
    doi = "10.22323/1.403.0002",
    journal = "PoS",
    volume = "TASI2021",
    pages = "002",
    year = "2023"
}

@article{Casini:2020,
    author = "Casini, Horacio and Huerta, Marina and Magan, Javier M. and Pontello, Diego",
    title = "{Entropic order parameters for the phases of QFT}",
    eprint = "2008.11748",
    archivePrefix = "arXiv",
    primaryClass = "hep-th",
    doi = "10.1007/JHEP04(2021)277",
    journal = "JHEP",
    volume = "04",
    pages = "277",
    year = "2021"
}

@article{Molina-Vilaplana:2024,
    author = "Molina-Vilaplana, Javier and Saura-Bastida, Pablo and Sierra, Germ{\'a}n",
    title = "{Entropic Order Parameters for Categorical Symmetries in 2D-CFT}",
    eprint = "2409.13460",
    archivePrefix = "arXiv",
    primaryClass = "hep-th",
    doi = "10.3390/e26121064",
    journal = "Entropy",
    volume = "26",
    number = "12",
    pages = "1064",
    year = "2024"
}

@article{AliAhmad:2025,
    author = "Ali Ahmad, Shadi and Klinger, Marc S. and Wang, Yifan",
    title = "{The many faces of non-invertible symmetries}",
    eprint = "2509.18072",
    archivePrefix = "arXiv",
    primaryClass = "hep-th",
    doi = "10.1007/JHEP05(2026)110",
    journal = "JHEP",
    volume = "05",
    pages = "110",
    year = "2026"
}

@article{Benini:2025lav,
    author = "Benini, Francesco and Calabrese, Pasquale and Fossati, Michele and Singh, Amartya Harsh and Venuti, Marco",
    title = "{Entanglement asymmetry for higher and noninvertible symmetries}",
    eprint = "2509.16311",
    archivePrefix = "arXiv",
    primaryClass = "hep-th",
    reportNumber = "SISSA 10/2025/FISI",
    month = "9",
    year = "2025"
}

@article{Zhang:2025,
    author = "Zhang, Hua-Chen and Sierra, Germ{\'a}n and Molina-Vilaplana, Javier",
    title = "{Entropic order parameters and topological holography}",
    eprint = "2512.24225",
    archivePrefix = "arXiv",
    primaryClass = "hep-th",
    doi = "10.1007/JHEP06(2026)083",
    journal = "JHEP",
    volume = "06",
    pages = "083",
    year = "2026"
}

@article{Araki2,
    author = "Araki, H.",
    title = "{Inequalities in von Neumann Algebras}",
    year = "1975"
}

@article{Araki:1976,
    author = "Araki, H.",
    title = "{Relative Entropy of States of Von Neumann Algebras}",
    journal = "Publ. Res. Inst. Math. Sci. Kyoto",
    volume = "1976",
    pages = "809--833",
    year = "1976"
}

@book{Petz:2004,
  title={Quantum Entropy and Its Use},
  author={Ohya, Masanori and Petz, D{\'e}nes},
  isbn={9783540208068},
  url={https://books.google.com/books/about/Quantum_Entropy_and_Its_Use.html?id=r2ullNVyESQC},
  year={2004},
  publisher={Springer-Verlag},
  address={Berlin, Heidelberg}
}

@book{Petz:2008,
  title     = {Quantum Information Theory and Quantum Statistics},
  author    = {Petz, Dénes},
  year      = {2008},
  publisher = {Springer Berlin, Heidelberg},
  address   = {Berlin, Heidelberg},
  isbn      = {978-3-540-74634-8},
  url       = {https://springer.com}
}

@article{Fiedler:2016,
    author = "Fiedler, Leander and Naaijkens, Pieter and Osborne, Tobias J.",
    title = "{Jones index, secret sharing and total quantum dimension}",
    eprint = "1608.02618",
    archivePrefix = "arXiv",
    primaryClass = "quant-ph",
    doi = "10.1088/1367-2630/aa5c0c",
    journal = "New J. Phys.",
    volume = "19",
    number = "2",
    pages = "023039",
    year = "2017"
}

@article{Naaijkens:2018,
    author = "Naaijkens, Pieter",
    title = "{Subfactors and quantum information theory}",
    eprint = "1704.05562",
    archivePrefix = "arXiv",
    primaryClass = "math-ph",
    doi = "10.1090/conm/717/14453",
    journal = "Contemp. Math.",
    volume = "717",
    pages = "257--279",
    year = "2018"
}

@article{Longo:1994xe,
    author = "Longo, R. and Rehren, Karl-Henning",
    title = "{Nets of subfactors}",
    eprint = "hep-th/9411077",
    archivePrefix = "arXiv",
    reportNumber = "DESY-94-205",
    doi = "10.1142/S0129055X95000232",
    journal = "Rev. Math. Phys.",
    volume = "7",
    pages = "567--598",
    year = "1995"
}

@article{Kong:2014,
       author = {{Kong}, Liang},
        title = "{Anyon condensation and tensor categories}",
      journal = {Nuclear Physics B},
         year = {2014},
        month = sep,
       volume = {886},
        pages = {436-482},
          doi = {10.1016/j.nuclphysb.2014.07.003},
archivePrefix = {arXiv},
       eprint = {1307.8244},
 primaryClass = {cond-mat.str-el},
       adsurl = {https://ui.adsabs.harvard.edu/abs/2014NuPhB.886..436K}
}

@book{Kodiyalam:2001,
  title     = {Topological Quantum Field Theories from Subfactors},
  author    = {Kodiyalam, Vijay and Sunder, V. S.},
  year      = {2001},
  publisher = {Chapman \& Hall/CRC},
  series    = {Research Notes in Mathematics},
  volume    = {423},
  isbn      = {9781584882411},
  url       = {https://www.routledge.com/Topological-Quantum-Field-Theories-from-Subfactors/Kodiyalam/p/book/9781138442108}
}

@article{Yu:2025,
    author = "Yu, Xingyang and Zhang, Hao Y.",
    title = "{Von Neumann subfactors and non-invertible symmetries}",
    eprint = "2504.05374",
    archivePrefix = "arXiv",
    primaryClass = "hep-th",
    doi = "10.21468/SciPostPhys.19.6.154",
    journal = "SciPost Phys.",
    volume = "19",
    number = "6",
    pages = "154",
    year = "2025"
}

@book{Watatani:1990,
  title={Index for $C^{*}$-subalgebras},
  author={Watatani, Yasuo},
  volume={83},
  number={424},
  series={Memoirs of the American Mathematical Society},
  year={1990},
  publisher={American Mathematical Society},
  address={Providence, RI},
  isbn={0-8218-2487-2}
}

@article{Jones:1983,
  title   = {Index for subfactors},
  author  = {Jones, Vaughan F. R.},
  journal = {Inventiones mathematicae},
  year    = {1983},
  volume  = {72},
  number  = {1},
  pages   = {1--25},
  doi     = {10.1007/BF01389127},
  issn    = {1432-1297}
}

@article{Bais1,
    author = "Bais, F. A. and Slingerland, J. K.",
    title = "{Condensate induced transitions between topologically ordered phases}",
    eprint = "0808.0627",
    archivePrefix = "arXiv",
    primaryClass = "cond-mat.mes-hall",
    reportNumber = "DIAS-STP-08-10, ITFA-2008-29",
    doi = "10.1103/PhysRevB.79.045316",
    journal = "Phys. Rev. B",
    volume = "79",
    pages = "045316",
    year = "2009"
}

@article{Bais2,
    author = {Eli{\"e}ns, I. S. and Romers, J. C. and Bais, F. A.},
    title = "{Diagrammatics for Bose condensation in anyon theories}",
    eprint = "1310.6001",
    archivePrefix = "arXiv",
    primaryClass = "cond-mat.str-el",
    doi = "10.1103/PhysRevB.90.195130",
    journal = "Phys. Rev. B",
    volume = "90",
    number = "19",
    pages = "195130",
    year = "2014"
}

@article{german2016,
  title={Boson condensation in topologically ordered quantum liquids},
  author={Titus Neupert and Huan He and C. W. von Keyserlingk and Germ{\'a}n Sierra and Andrei Bernevig},
  journal={Physical Review B},
  year={2016},
  volume={93},
  pages={115103},
  url={https://api.semanticscholar.org/CorpusID:118549636}
}

@article{Kitaev:2003,
    author = "Kitaev, Alexei and Mayers, Dominic and Preskill, John",
    title = "{Superselection rules and quantum protocols}",
    eprint = "quant-ph/0310088",
    archivePrefix = "arXiv",
    reportNumber = "CALT-68-2449, CALT-68-2449",
    doi = "10.1103/PhysRevA.69.052326",
    journal = "Phys. Rev. A",
    volume = "69",
    pages = "052326",
    year = "2004"
}

@article{Bonderson:2008,
    author = "Bonderson, Parsa and Freedman, Michael and Nayak, Chetan",
    title = "{Measurement-only topological quantum computation via anyonic interferometry}",
    eprint = "0808.1933",
    archivePrefix = "arXiv",
    primaryClass = "quant-ph",
    doi = "10.1016/j.aop.2008.09.009",
    journal = "Annals Phys.",
    volume = "324",
    pages = "787--826",
    year = "2009"
}

@article{Benedetti:2024dku,
    author = "Benedetti, Valentin and Casini, Horacio and Kawahigashi, Yasuyuki and Longo, Roberto and Magan, Javier M.",
    title = "{Modular invariance as completeness}",
    eprint = "2408.04011",
    archivePrefix = "arXiv",
    primaryClass = "hep-th",
    doi = "10.1103/PhysRevD.110.125004",
    journal = "Phys. Rev. D",
    volume = "110",
    number = "12",
    pages = "125004",
    year = "2024"
}

@article{Bockenhauer1,
    author = "Bockenhauer, Jens and Evans, David E.",
    title = "{On alpha induction, chiral generators and modular invariants for subfactors}",
    eprint = "math/9904109",
    archivePrefix = "arXiv",
    doi = "10.1007/s002200050765",
    journal = "Commun. Math. Phys.",
    volume = "208",
    pages = "429--487",
    year = "1999"
}

@article{Bockenhauer2,
    author = "Bockenhauer, Jens and Evans, David E. and Kawahigashi, Yasuyuki",
    title = "{Chiral structure of modular invariants for subfactors}",
    eprint = "math/9907149",
    archivePrefix = "arXiv",
    doi = "10.1007/s002200050798",
    journal = "Commun. Math. Phys.",
    volume = "210",
    pages = "733--784",
    year = "2000"
}

@article{Schafer-Nameki:2025,
    author = "Schafer-Nameki, Sakura and Tiwari, Apoorv and Warman, Alison and Zhang, Carolyn",
    title = "{SymTFT Approach for Mixed States with Non-Invertible Symmetries}",
    eprint = "2507.05350",
    archivePrefix = "arXiv",
    primaryClass = "quant-ph",
    month = "7",
    year = "2025"
}

@article{Kitaev:1997,
    author = "Kitaev, A. Yu.",
    title = "{Fault tolerant quantum computation by anyons}",
    eprint = "quant-ph/9707021",
    archivePrefix = "arXiv",
    doi = "10.1016/S0003-4916(02)00018-0",
    journal = "Annals Phys.",
    volume = "303",
    pages = "2--30",
    year = "2003"
}

@misc{Mueger:2001,
      title={From Subfactors to Categories and Topology I. Frobenius algebras in and Morita equivalence of tensor categories}, 
      author={Michael Mueger},
      year={2002},
      eprint={math/0111204},
      archivePrefix={arXiv},
      primaryClass={math.CT},
      url={https://arxiv.org/abs/math/0111204}, 
}

@article{Bhardwaj:2024,
    author = "Bhardwaj, Lakshya and Bottini, Lea E. and Schafer-Nameki, Sakura and Tiwari, Apoorv",
    title = "{Illustrating the categorical Landau paradigm in lattice models}",
    eprint = "2405.05302",
    archivePrefix = "arXiv",
    primaryClass = "cond-mat.str-el",
    doi = "10.1103/PhysRevB.111.054432",
    journal = "Phys. Rev. B",
    volume = "111",
    number = "5",
    pages = "054432",
    year = "2025"
}

@article{Eck:2023gic,
    author = "Eck, Luisa and Fendley, Paul",
    title = "{From the XXZ chain to the integrable Rydberg-blockade ladder via non-invertible duality defects}",
    eprint = "2302.14081",
    archivePrefix = "arXiv",
    primaryClass = "cond-mat.stat-mech",
    doi = "10.21468/SciPostPhys.16.5.127",
    journal = "SciPost Phys.",
    volume = "16",
    number = "5",
    pages = "127",
    year = "2024"
}

@article{Teruya1992,
  author  = {Teruya, Tamotsu},
  title   = {Index for von {N}eumann algebras with
             finite-dimensional centers},
  journal = {Publ. Res. Inst. Math. Sci.},
  volume  = {28},
  pages   = {437--453},
  year    = {1992}
}

@article{Giorgetti_2018,
   title={Minimal Index and Dimension for 2-C*-Categories with 
   Finite-Dimensional Centers},
   volume={370},
   ISSN={1432-0916},
   url={http://dx.doi.org/10.1007/s00220-018-3266-x},
   DOI={10.1007/s00220-018-3266-x},
   number={2},
   journal={Communications in Mathematical Physics},
   publisher={Springer Science and Business Media LLC},
   author={Giorgetti, Luca and Longo, Roberto},
   year={2018},
   month=Oct, pages={719–757} }

@article{Magan:2020ake,
    author = "Magan, Javier M. and Pontello, Diego",
    title = "{Quantum Complementarity through Entropic Certainty Principles}",
    eprint = "2005.01760",
    archivePrefix = "arXiv",
    primaryClass = "hep-th",
    doi = "10.1103/PhysRevA.103.012211",
    journal = "Phys. Rev. A",
    volume = "103",
    number = "1",
    pages = "012211",
    year = "2021"
}

@article{Benedetti:2026,
    author = "Benedetti, Valentin and Fendley, Paul and Magan, Javier M.",
    title = "{Non-invertible symmetries and selection rules for RG flows of coset models}",
    eprint = "2603.09591",
    archivePrefix = "arXiv",
    primaryClass = "hep-th",
    month = "3",
    year = "2026"
}

@article{Bais:2002,
  author  = {Bais, F. A. and Schroers, B. J. and Slingerland, J. K.},
  title   = {Broken quantum symmetry and confinement phases in planar physics},
  journal = {Phys. Rev. Lett.}, volume = {89}, pages = {181601}, year = {2002}
}

@article{Burnell:2011,
  author  = {Burnell, F. J. and Simon, S. H. and Slingerland, J. K.},
  title   = {Condensation of achiral simple currents in topological lattice
             models: {H}amiltonian study of topological symmetry breaking},
  journal = {Phys. Rev. B}, volume = {84}, pages = {125434}, year = {2011}
}

@article{Burnell:2012,
  author  = {Burnell, F. J. and Simon, S. H. and Slingerland, J. K.},
  title   = {Phase transitions in topological lattice models via topological
             symmetry breaking},
  journal = {New J. Phys.}, volume = {14}, pages = {015004}, year = {2012}
}


\end{document}